\documentclass[12pt]{iopart} 
\usepackage[dvips]{graphicx} 
 \usepackage{iopams}
\begin{document}

\title{Superfluid-density of the ultra-cold Fermi gas in optical lattices}
\author{T. Paananen$^{1,2}$}
\address{$^1$ Department of Physics, P.O. Box 64, FI-00014 University of Helsinki,  Finland}
\address{$^2$ Laboratory of Physics, Helsinki University of Technology, P.O. Box 1100, 02015 HUT, Finland}
\ead{tomi.paananen@helsinki.fi}

\date{\today}
\begin{abstract}
In this paper we study the superfluid density of the two component Fermi gas 
in optical lattices with population imbalance.
Three different type of phases, the BCS-state (Bardeen, Cooper, and Schrieffer), 
the FFLO-state  
(Fulde, Ferrel, Larkin, and Ovchinnikov), and the Sarma state, are considered.
We show that the FFLO superfluid density differs from the BCS/Sarma
superfluid density in an important way. 
Although there are dynamical instabilities in the FFLO phase, when
the interaction is strong or densities are high, on the weak
coupling limit the FFLO phase is found to be stable.

\end{abstract}
\submitto{\JPB} 
\maketitle

\section{Introduction}
\label{sec:intro}

The field of ultra-cold Fermi gases offers a great tool to study many different problems
of correlated quantum systems.
For example, in recent
experiments~\cite{Zwierlein2006a,Partridge2006a,Zwierlein2006c,Shin2006a,Partridge2006c,Shin2008a} 
polarized Fermi gases were considered. 
These systems make it possible to study physics in the presence of mismatched Fermi surfaces, 
and non-BCS type pairing 
such as that appearing in FFLO-states~\cite{Fulde1964a,Larkin1964a} 
or Sarma-states~\cite{Sarma1963a,Liu2003a}. These possibilities have been considered 
extensively in condensed-matter, nuclear, and high-energy physics~\cite{Casalbuoni2004a}.  

It is also possible to study  many different physical problems with close analogs 
in the field of solid state physics using optical lattices. However, unlike solid state systems, 
ultra-cold gases in optical lattices provide a very clean environment.  These systems have very 
few imperfections and if one is interested in imperfections, they can be imposed easily on the system. 
Optical lattices are made with lasers, thus the lattice geometry is easy to 
modify~\cite{Orzel2001a,Greiner2001b,Burger2001a,Hadzibabic2004a} by changing
the properties of the intersecting laser beams.
For these reasons, in optical lattices one can study   
various quantum many-body physics problems, such as 
Mott insulators, phase coherence, and superfluidity.
The possibility of a superfluid alkali atom Fermi gas in an optical lattice 
has been recently studied both 
theoretically~\cite{Hofstetter2002a,Orso2005a,Pitaevskii2005a,Iskin2007a,Koponen2006a,Koponen2006b,Koponen2007a}, 
as well as experimentally~\cite{Chin2006a}.

The pairing predicted by the BCS theory does not always mean that the gas is a superfluid~\cite{Schunck2007a}. 
When a linear phase is imposed on the order parameter, the phase gradient corresponds to superfluid velocity, and a part of the gas, which has superfluid velocity is
called superfluid, and the coefficient of the inertia of this moving part is called superfluid density~\cite{Fisher1973a,Rey2003a,Roth2003a}.
However, if the gas is normal the order parameter vanishes and thus the phase shift 
does not affect to the normal gas.
If the superfluid density is positive in all directions the gas is a superfluid.
Negative superfluid density implies dynamical instability of the gas~\cite{Wu2001a,Burkov2008a}.

There are a fair number of studies about superfluid density of Fermi gas in the free space
as well as in a trap~\cite{Taylor2006a,Fukushima2007a,Taylor2008a,Stojanovic2008a},
and few studies about superfluid density in optical lattices~\cite{Iskin2005a,Burkov2008a}.
However, these papers have focused on population balanced cases i.e. the densities of the components are same.
In this paper we study  the superfluid density of an ultra-cold two component Fermi gas in optical lattices
at finite temperatures and with finite polarization.
We are motivated by the fact by calculating the superfluid density we can draw
more conclusions on whether the gas is actually superfluid or not. 
Furthermore, we investigate if the superfluid density can be used as
an indicator of different superfluid phases.
The phases we consider are the BCS phase,i.e. the densities of the component are equal,
the one mode FFLO phase, i.e. the densities are not equal and the phase of the pairing
gap modulates as a function of position,  and Sarma phase i.e.  the densities are not equal and the phase of the pairing
gap is constant.
We show that there are qualitatively difference between the BCS/Sarma superfluid density
and the one mode FFLO superfluid density.
We also study the stability of different phases.
We also demonstrate that there can be dynamical instabilities in the one mode FFLO phase.

This paper is organized as follows. In Sec.~\ref{sec:hub_mod}
we discuss the physical system and present the Hamiltonian of the system. 
In Subsec.~\ref{subsec:meanfield} 
the mean-field approximation and the ansatzes we consider are presented.
In Sec.~\ref{sec:super_dens} we determine the superfluid density tensor.
In Sec.~\ref{sec:results} we present the numerical results and
we end with some concluding remarks in Sec.~\ref{sec:conc}.

\section{Lowest band fermions in an optical lattice}
\label{sec:hub_mod}

We consider a two component Fermi gas, whose  components are two different hyperfine states of the same isotope, and we call them $\uparrow$-state and $\downarrow$-state.
The Hamiltonian of the system is given by
\begin{eqnarray}
\label{eq:tot_Hamltonian}
&\hat H=\sum_{\sigma=\uparrow,\downarrow}\int\,d{\bf r}\,\hat \Psi^{\dagger}_{\sigma}({\bf r})\left(-\frac{\hbar^2 \nabla^2}{2m}+V_{\sigma}({\bf r})-\mu_{\sigma}\right)\hat
 \Psi_{\sigma}({\bf r})\\ \nonumber
&+g\int\,\int\,d{\bf r}\,d{\bf r'}\hat \Psi_{1}^{\dagger}({\bf r})\hat \Psi_{2}^{\dagger}({\bf r'})\delta({\bf r}-{\bf r'})\hat \Psi_{2}({\bf r'})\hat \Psi_{1}({\bf r})
\end{eqnarray}
where $\hbar=h/2\pi$, $h$ is Planck's constant, $\Psi_{\sigma}^{\dagger}({\bf r})$ and $\Psi_{\sigma}({\bf r})$ are the fermionic 
 creation- and annihilation field operators of the component $\sigma$, $\mu_{\sigma}$ is the
 chemical potential of the component $\sigma$, and $\delta({\bf r}-{\bf r'})$ is  Dirac's $\delta$-function.
The interaction strength is related to the s-wave scattering length $a_s$ through
\[ g=\frac{4\pi\hbar^2 a_s}{m}.\]
The lattice potential has a cubic structure and is given by
\[ V_{\sigma}({\bf r})=E_r\sum_{\alpha}s_{\sigma,\alpha}\sin^2(kx_\alpha),\]
where $E_r=\hbar^2k^2/2m$ is the recoil energy ($k=(\pi/d)$ where $d$ is a lattice constant), and $s_{\sigma,\alpha}$ is the lattice depth in the  $\alpha$-direction for 
the component $\sigma$.

When we assume that only the lowest band is occupied the field operators can be expanded by using the localized Wannier functions in the following way
\begin{eqnarray}
&\hat \Psi_{\sigma}({\bf r})=\sum_{i}w_{\sigma,i}({\bf r})\hat c_{\sigma,i}\\
&\hat \Psi^{\dagger}_{\sigma}({\bf r})=\sum_{i}w_{\sigma,i}^*({\bf r})\hat c^{\dagger}_{\sigma,i},
\end{eqnarray}
where $w_{\sigma,i}({\bf r})$ is a Wannier function, which is localized at a lattice point $i$, and $\hat c_{\sigma,i}$ is an annihilation operator.

The lowest band Hubbard model is valid, when the lattice is deep enough.
In other words, it is valid, when the Wannier functions decay within a single lattice constant, and
when the effective interaction between atoms is much smaller than the bandgap between bands.
These two conditions imply that one has to take into account only the on-site interactions.
In this case overlap integrals of the kinetic energy operator between the next nearest neighbor are
small compared to overlap integrals of the kinetic energy operator between the nearest neighbor~\cite{Bloch2007a},
and consequently, one needs to take into account only hopping between the nearest neighbours.
Then the lowest band Hubbard Hamiltonian is given by 
\begin{eqnarray}
\label{eq:hub_ham}
&\hat H=-\sum_{\sigma=\uparrow,\downarrow}\left(J_{\sigma,x}\sum_{\langle i,j\rangle_x}+J_{\sigma,y}\sum_{\langle i,j\rangle_y}+J_{\sigma,z}\sum_{\langle i,j\rangle_z}\right)
\hat c^{\dagger}_{\sigma,i}\hat c_{\sigma,j}\\ \nonumber
&-\sum_i(\mu_{\uparrow} c^{\dagger}_{\uparrow,i}\hat c_{\uparrow,i}+\mu_{\downarrow} c^{\dagger}_{\downarrow,i}\hat c_{\downarrow,i})
+U\sum_i\hat c^{\dagger}_{\uparrow,i}\hat c^{\dagger}_{\downarrow,i}\hat c_{\downarrow,i}\hat c_{\uparrow,i},
\end{eqnarray}
where $\langle i,j\rangle_{\alpha}$ means sum over the nearest neighbours in the $\alpha$-direction,
and hopping strength is defined by 
\[
J_{\sigma,\alpha}=-\int\, d{\bf r}\, w^*_{\sigma,i}({\bf r})\left(-\frac{\hbar^2\nabla^2}{2m}+V_{\sigma}({\bf r})\right)w_{\sigma,i\pm d\hat x_{\alpha}}({\bf r}).
\]
The coupling strength of the above Hubbard model is given by
\[
U=g\int\,d{\bf r}\,|w_{\uparrow,i}({\bf r})|^2|w_{\downarrow,i}({\bf r})|^2.
\]

\subsection{Mean-field approximation}
\label{subsec:meanfield}

Because the interaction term $U\sum_i\hat c^{\dagger}_{\uparrow,i}\hat c^{\dagger}_{\downarrow,i}\hat c_{\downarrow,i}\hat c_{\uparrow,i}$ is hard to handle,
we can approximate it by using mean-field approximation. Under this approximation the interaction part becomes
\[ \sum_i\Delta(i)\hat c_{\uparrow,i}^{\dagger}\hat c_{\downarrow,i}^{\dagger}+\Delta^*(i)\hat c_{\downarrow,i}\hat c_{\uparrow,i}-\frac{|\Delta(i)|^2}{U},\]
where the pairing gap $\Delta(i)=U\langle\hat c_{\downarrow,i}\hat c_{\uparrow,i}\rangle$.
With respect to position dependence, we only consider the case, in which only the phase of the gap $\Delta(i)=|\Delta| e^{i{\bf q}\cdot {\bf R}_i}$
(${\bf R}_i$ is a lattice vector) can depend on the lattice site $i$. 
This ansatz is called the one mode FFLO (Fulde, Ferrell, Larkin, and Ovchinnikov)~\cite{Fulde1964a,Larkin1964a}. This ansatz includes also the BCS-ansatz and
the Sarma-state ansatz as special cases. In these phases the momentum ${\bf q}$ is simply zero.
Under this mean.field approximation the Hamiltonian is given by
\begin{eqnarray}
\label{eq:meanfieldham}
&\hat H_0=-\sum_{\sigma=\uparrow,\downarrow}\left(J_{\sigma,x}\sum_{\langle i,j\rangle_x}+J_{\sigma,y}\sum_{\langle i,j\rangle_y}+J_{\sigma,z}\sum_{\langle i,j\rangle_z}\right)
\hat c^{\dagger}_{\sigma,i}\hat c_{\sigma,j}\\ \nonumber
&-\sum_i(\mu_{\uparrow} c^{\dagger}_{\uparrow,i}\hat c_{\uparrow,i}+\mu_{\downarrow} c^{\dagger}_{\downarrow,i}\hat c_{\downarrow,i})
+\sum_i |\Delta| e^{i{\bf q}\cdot {\bf R}_i}\hat c^{\dagger}_{\uparrow,i}\hat c^{\dagger}_{\downarrow,i}+|\Delta| e^{-i{\bf q}\cdot {\bf R}_i}\hat c_{\downarrow,i}\hat c_{\uparrow,i}.
\end{eqnarray}

By minimizing the free energy $F_0=\Omega_0+\mu_{\uparrow}N_{\uparrow}+\mu_{\downarrow}N_{\downarrow}$ ($\Omega_0$ is the grand canonical
potential of the mean-field Hamiltonian) with respect $|\Delta|$ and ${\bf q}$, and solving the number equations 
$\partial F_0/\partial \mu_{\sigma}=0$ simultaneously,
one finds the pairing gap, the chemical potentials, and the momentum ${\bf q}$ as functions of the temperature and the particle numbers $N_{\sigma}$.

\section{Superfluid density}
\label{sec:super_dens}

Landau's two component model for a superfluid gas says that the superfluid gas consists two components the normal component
and the superfluid component.  In the free space the superfluid density is density of the superfluid component.
The case is a little bit different in lattices thus the energy difference between the twisted system and the system without
the twisting phase can depend on the direction of the phase shift. This is related the fact that 
the effective masses can depend on the direction  (the effective masses are related to the hopping strengths). Thus
in the lattice superfluid density is more like a tensor than a scalar.

In the free space when the superfluid gas flows the kinetic energy of the gas
is given by
\[
E_k=\frac{1}{2}\int\, d{\bf r}\, \tilde \rho_s({\bf r}) v_s({\bf r})^2,
\]
where $\tilde \rho_s({\bf r})$ is the superfluid density and $v_s({\bf r})$ is the superfluid velocity. 
The superfluid velocity is defined by
\[{\bf v}_s({\bf r})=\frac{\hbar}{2m}\nabla \phi({\bf r}),\]
where $\phi({\bf r})$ is the phase of the order parameter.

To impose the linear phase variation to the order parameter,  one can impose a linear phase variation 
${\bf \Theta}\cdot{\bf R}_i=(\Theta_x/(M_xd),\Theta_y/(M_yd),\Theta_z/(M_zd))\cdot{\bf R}_i$~\cite{Rey2003a,Roth2003a, Martikainen2004b} to the Hamiltonian.
Here $M_{\alpha}$ indicates the number of lattice sites in direction $alpha$, i.e., the total volume the lattice $V=M_xM_yM_zd^3$
This variation corresponds small ($\Theta_{\alpha}$ is small) superfluid velocity, which is given by
\[
{\bf v}_s=\frac{\hbar}{2m}(\Theta_x/(M_xd),\Theta_y/(M_yd),\Theta_z/(M_zd)).
\]
Thus this imposed phase gradient gives the system a kinetic energy, which corresponds
to the free energy difference $F_{\bf \Theta}-F_0$, where $F_{\bf \Theta}$ is the free energy within the  the phase variation and
$F_0$ is the free energy without the phase variation.
The superfluid fraction in this case can be determined as~\cite{Rey2003a,Roth2003a}
\begin{equation}
\rho_{\alpha\alpha'}=\lim_{{\bf \Theta}\rightarrow 0}\frac{1}{N}\frac{F_{\bf \Theta}-F_0}{\bar{J_x}\Theta_{\alpha}\Theta_{\alpha'}}=
\frac{1}{N\bar{J_x}}\frac{\partial^2 F_{\bf \Theta}}{\partial \Theta_{\alpha} \partial \Theta_{\alpha'}}\bigg |_{{\bf \Theta}=0},
\end{equation}
where $N$ is the total number of particles, and $\bar{J_x}=(J_{\uparrow,x}+J_{\downarrow,x})/2$. This $\bar{J_x}$ corresponds
the effective mass \[m_{eff}=\frac{\hbar^2}{2\bar{J_x}d^2}.\]
When one takes the limit ${\bf \Theta}\rightarrow 0$ the temperature, and
the number of particles (and of course the interaction strength) should be kept as constants.

 Of course, one could define the superfluid fraction as
\begin{equation}
\label{eq:2_maar}
\rho_{\alpha\alpha}=\lim_{{\bf \Theta}\rightarrow 0}\frac{1}{N}\frac{F_{\bf \Theta}-F_0}{\bar{J_{\alpha,\alpha}}\Theta_{\alpha}\Theta_{\alpha'}}
\end{equation}
where $\bar{J_{\alpha,\alpha'}}=(J_{\uparrow,\alpha}+J_{\downarrow,\alpha})/2$, but in this case it is a little bit unclear how 
to define the off-diagonal elements. This definition is not directly connected to the energy difference between the twisted
system and the untwisted system.

By imposing the phase variation to the order parameter the one finds
\begin{eqnarray}
\label{eq:twistHam_1}
&\hat H_{\bf \Theta}=-\sum_{\sigma=\uparrow,\downarrow}\left(J_{\sigma,x}\sum_{\langle i,j\rangle_x}+J_{\sigma,y}\sum_{\langle i,j\rangle_y}+J_{\sigma,z}\sum_{\langle i,j\rangle_z}\right)
\hat c^{\dagger}_{\sigma,i}\hat c_{\sigma,j}\\ \nonumber
&-\sum_i(\mu_{\uparrow} c^{\dagger}_{\uparrow,i}\hat c_{\uparrow,i}+\mu_{\downarrow} c^{\dagger}_{\downarrow,i}\hat c_{\downarrow,i})
+\sum_i |\Delta| e^{i({\bf q}+2{\bf \Theta})\cdot {\bf R}_i}\hat c^{\dagger}_{\uparrow,i}\hat c^{\dagger}_{\downarrow,i}+|\Delta| e^{-i({\bf q}+2{\bf \Theta})
\cdot {\bf R}_i}\hat c_{\downarrow,i}\hat c_{\uparrow,i}.
\end{eqnarray}
One can make the unitary transformation~\cite{Rey2003a,Roth2003a,Martikainen2004b}, 
which corresponds the local transformation $\hat c_{\sigma,i}\rightarrow \hat c_{\sigma,i}e^{i{\bf \Theta}\cdot {\bf R}_i}$,
and the twisted Hamiltonian becomes
\begin{eqnarray}
\label{eq:twistHam_2}
&\hat H_{\bf \Theta}=-\sum_i(\mu_{\uparrow} c^{\dagger}_{\uparrow,i}\hat c_{\uparrow,i}+\mu_{\downarrow} c^{\dagger}_{\downarrow,i}\hat c_{\downarrow,i})\\ \nonumber
&+\sum_i |\Delta| e^{i{\bf q}\cdot {\bf R}_i}\hat c^{\dagger}_{\uparrow,i}\hat c^{\dagger}_{\downarrow,i}+|\Delta| e^{-i{\bf q}
\cdot {\bf R}_i}\hat c_{\downarrow,i}\hat c_{\uparrow,i}\\ \nonumber
&-\sum_{\sigma}\bigg[\sum_n\bigg(J_{\sigma,x}(e^{i\Theta_x/M_x}\hat c^{\dagger}_{\sigma,n}\hat c_{\sigma,n+d\hat x}+e^{-i\Theta_x/M_x}\hat c^{\dagger}_{\sigma,n}\hat c_{\sigma,n-d\hat x})\\ 
\nonumber
&+J_{\sigma,y}(e^{i\Theta_y/M_y}\hat c^{\dagger}_{\sigma,n}\hat c_{\sigma,n+d\hat y}+e^{-i\Theta_y/M_y}\hat c^{\dagger}_{\sigma,n}\hat c_{\sigma,n-d\hat y})\\ \nonumber
&+J_{\sigma,z}(e^{i\Theta_z/M_z}\hat c^{\dagger}_{\sigma,n}\hat c_{\sigma,n+d\hat z}+e^{-i\Theta_z/M_z}\hat c^{\dagger}_{\sigma,n}\hat c_{\sigma,n-d\hat z})\bigg)\bigg].
\end{eqnarray}
The twist angles $\Theta_{\alpha}$ have to be sufficiently small to avoid effects other than the collective flow of the superfluid component.
Since $\Theta_{\alpha}/M_{\alpha}$ is small, we can expand up to second order
\[e^{\pm i\Theta_{\alpha}/M_{\alpha}}\approx 1\pm i\frac{\Theta_{\alpha}}{M_{\alpha}}-\frac{\Theta_{\alpha}^2}{2M_{\alpha}^2}.\] 
In this way we can write the twisted Hamiltonian as
\begin{eqnarray}
\label{eq:twistHam_3}
&\hat H_{\bf \Theta}=\hat H_0+H'\approx\hat H_0+\sum_{\sigma}\bigg[\frac{\Theta_x^2}{2M_x^2}\sum_{n}J_{\sigma,x}(\hat c^{\dagger}_{\sigma,n}\hat c_{\sigma,n+d\hat x}+
\hat c^{\dagger}_{\sigma,n}\hat c_{\sigma,n-d\hat x})+\\ \nonumber
&\frac{\Theta_y^2}{2M_y^2}\sum_{n}J_{\sigma,y}(\hat c^{\dagger}_{\sigma,n}\hat c_{\sigma,n+d\hat y}+\hat c^{\dagger}_{\sigma,n}\hat c_{\sigma,n-d\hat y})
+\frac{\Theta_z^2}{2M_z^2}\sum_{n}J_{\sigma,z}(\hat c^{\dagger}_{\sigma,n}\hat c_{\sigma,n+d\hat z}+\hat c^{\dagger}_{\sigma,n}\hat c_{\sigma,n-d\hat z})\\ \nonumber
&-i\frac{\Theta_x}{M_x}\sum_nJ_{\sigma,x}(\hat c^{\dagger}_{\sigma,n}\hat c_{\sigma,i+d\hat x}-\hat c^{\dagger}_{\sigma,n}\hat c_{\sigma,n-d\hat x})-
i\frac{\Theta_y}{M_y}\sum_nJ_{\sigma,y}(\hat c^{\dagger}_{\sigma,n}\hat c_{\sigma,i+d\hat y}-\hat c^{\dagger}_{\sigma,n}\hat c_{\sigma,n-d\hat y})\\ \nonumber
&-i\frac{\Theta_z}{M_z}\sum_nJ_{\sigma,z}(\hat c^{\dagger}_{\sigma,n}\hat c_{\sigma,i+d\hat z}-\hat c^{\dagger}_{\sigma,n}\hat c_{\sigma,n-d\hat z})\bigg]\nonumber.
\end{eqnarray} 
In the momentum space this formula becomes
\begin{eqnarray}
\label{eq:twistHam_4}
\hat H_{\bf \Theta}&\approx\hat H_0+\sum_{\sigma,\alpha}\bigg[\frac{\Theta_{\alpha}^2}{2M_{\alpha}^2}\sum_{\bf k}2J_{\sigma,\alpha}\cos(k_{\alpha}d)
\hat c^{\dagger}_{\sigma,{\bf k}}\hat c_{\sigma,{\bf k}}\\ \nonumber
&+\frac{\Theta_{\alpha}}{M_{\alpha}}
\sum_{\bf k}2J_{\sigma,\alpha}\sin(k_{\alpha}d) \hat c^{\dagger}_{\sigma,{\bf k}}\hat c_{\sigma,{\bf k}}\bigg]\\ \nonumber
&=\hat H_0+\hat T+\hat J, \nonumber
\end{eqnarray}
where 
\begin{eqnarray}
\hat T=\sum_{\sigma,\alpha}\bigg[\frac{\Theta_{\alpha}^2}{2M_{\alpha}^2}\sum_{\bf k}2J_{\sigma,\alpha}\cos(k_{\alpha}d)
\hat c^{\dagger}_{\sigma,{\bf k}}\hat c_{\sigma,{\bf k}}\bigg]
\end{eqnarray}
and
\begin{eqnarray}
\hat J=\sum_{\sigma,\alpha}\bigg[\frac{\Theta_{\alpha}}{M_{\alpha}}
\sum_{\bf k}2J_{\sigma,\alpha}\sin(k_{\alpha}d) \hat c^{\dagger}_{\sigma,{\bf k}}\hat c_{\sigma,{\bf k}}\bigg].
\end{eqnarray}
These two terms are proportional to the current, and number operator of the system, respectively.
These two terms commute with the number operators, thus when one imposes the perturbation in the system  the number of particles is conserved.
Using the same method, what is used in reference~\cite{Taylor2006a},  can be shown that in the mean-field level
\begin{equation}
\lim_{|{\bf \Theta}|\rightarrow 0}\frac{\partial^2 F({\bf \Theta})}{\partial \Theta_{\alpha}\partial \Theta_{\alpha'}}
=\lim_{|{\bf \Theta}|\rightarrow 0}\left(\frac{\partial^2 \Omega({\bf \Theta})}{\partial \Theta_{\alpha}\partial \Theta_{\alpha'}}\right)_{\Delta,\mu_{\uparrow},\mu_{\downarrow}}.
\end{equation}
Therefore we can use the grand canonical potential.

The grand potential of the system can be written with a perturbing phase gradient as a series~\cite{MahanMNP}
\begin{equation}
\label{eq:twist_grandpot}
\Omega_{\bf \Theta}=\Omega_0-\sum^{\infty}_{n=1}\frac{(-1)^n}{\beta n\hbar^n}\int_0^{\beta\hbar}d\tau_1\,\cdots\int_0^{\beta\hbar}d\tau_n\,\langle T_{\tau}\hat H'(\tau_1)\cdots
\hat H'(\tau_n)\rangle_0,
\end{equation}
where $\beta=1/k_BT$, $k_B$ is Boltzmann's constant, $\Omega_0$ is the grand canonical potential in the absence of the perturbation.
The symbol $T_{\tau}$ orders the operators such a way that $\tau$ decreases from left to right.
The brackets $\langle\dots\rangle_0 = \Tr[\exp(-\beta(\hat H_0-\mu_{\uparrow}\hat N_{\uparrow}-\mu_{\downarrow}\hat N_{\downarrow} ))\dots]$ mean the thermodynamic 
average evaluated in the equilibrium state of the unperturbed system at temperature $T$. Because all the twisted angles $\Theta_{\alpha}$ are small,
we can safely ignore terms of higher order than $\Theta_{\alpha}^2$. 
With this approximation the grand potential becomes
\begin{equation}
\label{eq:twist_grandpot2}
\Omega_{\bf \Theta}\approx\Omega_0+\frac{1}{\beta \hbar}\int_0^{\beta\hbar}d\tau\,\langle T_{\tau}\hat T(\tau)\rangle_0-
\frac{1}{2\beta \hbar^2}\int_0^{\beta\hbar}\int_0^{\beta\hbar}d\tau\,d\tau'\langle T_{\tau}\hat J(\tau)J(\tau')\rangle_0,
\end{equation}
The first perturbation term is easy to calculate and it is given by
\begin{equation}
\label{eq:first_pertuterm}
\frac{1}{\beta \hbar}\int_0^{\beta\hbar}d\tau\,\langle T_{\tau}\hat T(\tau)\rangle_0=
\sum_{\sigma,\alpha}\bigg[\frac{\Theta_{\alpha}^2}{2M_{\alpha}^2}\sum_{\bf k}2J_{\sigma,\alpha}\cos(k_{\alpha}d)N_{\sigma,{\bf k}}\bigg],
\end{equation}
where $N_{\sigma,{\bf k}}$ is number of particles of the $\sigma$-component in the momentum state ${\bf k}$.
Using Wick's theorem the second perturbation term is given by
\begin{eqnarray}
&-\frac{1}{2\beta \hbar^2}\int_0^{\beta\hbar}\int_0^{\beta\hbar}\,d\tau\,d\tau'\,\langle T_{\tau}\hat J(\tau)\hat J(\tau')\rangle_0\\ \nonumber
&=-\frac{1}{2\beta \hbar^2}\int_0^{\beta\hbar}\int_0^{\beta\hbar}\,d\tau\,d\tau'\,\bigg[\sum_{\sigma,\sigma',\alpha,\alpha'}
\frac{\Theta_{\alpha}\Theta_{\alpha'}}{M_{\alpha}M_{\alpha'}}\sum_{{\bf k},{\bf k'}}4J_{\sigma,\alpha}J_{\sigma',\alpha'}\sin(k_{\alpha}d)\sin(k'_{\alpha'}d)\\ \nonumber
&\times\bigg((1-\delta_{\sigma\sigma'})\delta_{{\bf k},-{\bf k'}+{\bf q}}F({\bf k},\tau,\tau')F^{\dagger}({\bf k'},\tau,\tau')-\delta_{\sigma\sigma'}\delta_{{\bf k}{\bf k'}}
G_{\sigma}({\bf k},\tau,\tau')G_{\sigma'}({\bf k'},\tau,\tau')\bigg)\bigg],
\end{eqnarray}
where 
\begin{eqnarray}
&G_{\sigma}({\bf k},\tau,\tau')=-\langle T_{\tau}\hat c_{\sigma,{\bf k}}(\tau)\hat c^{\dagger}_{\sigma,{\bf k}}(\tau') \rangle_0=
\frac{1}{\sqrt{\beta\hbar}}\sum_{n=-\infty}^{\infty}e^{-i\omega_n(\tau-\tau')}G_{\sigma}({\bf k},\omega_n)\\
&F({\bf k},\tau,\tau')=-\langle T_{\tau}\hat c_{\uparrow,{\bf k}+{\bf q}}(\tau)\hat c_{\downarrow,-{\bf k}+{\bf q}}(\tau') \rangle_0=
\frac{1}{\sqrt{\beta\hbar}}\sum_{n=-\infty}^{\infty}e^{-i\omega_n(\tau-\tau')}F({\bf k},\omega_n).\nonumber
\end{eqnarray} 
The fermionic Matsubara frequencies are given by
\[
\omega_n=\frac{\pi(2n+1)}{\hbar\beta},
\]
where $n$ is an integer.
The individual Green's functions of the mean-field theory can be written as
\begin{eqnarray}
&G_{\uparrow}({\bf k},\omega_n)=\frac{|u_{{\bf k},{\bf q}}|^2}{i\hbar\omega_n-E_{+,{\bf k},{\bf q}}}+\frac{|v_{{\bf k},{\bf q}}|^2}{i\hbar\omega_n+E_{-,{\bf k},{\bf q}}}\\ 
&G_{\downarrow}({\bf k},\omega_n)= \frac{|u_{{\bf k},{\bf q}}|^2}{i\hbar\omega_n-E_{-,{\bf k},{\bf q}}}+\frac{|v_{{\bf k},{\bf q}}|^2}{i\hbar\omega_n+E_{+,{\bf k},{\bf q}}}\\
&F({\bf k},\omega_n)= \frac{u_{{\bf k},{\bf q}}v^*_{{\bf k},{\bf q}}}{i\hbar\omega_n+E_{-,{\bf k},{\bf q}}}-
\frac{u_{{\bf k},{\bf q}}v^*_{{\bf k},{\bf q}}}{i\hbar\omega_n-E_{+,{\bf k},{\bf q}}}, 
\end{eqnarray}
where the quasiparticle dispersions and amplitudes are given by
\begin{eqnarray}
&E_{\pm,{\bf k},{\bf q}}=\sqrt{\left(\frac{\epsilon_{\uparrow,{\bf k}+{\bf q}/2}+\epsilon_{\downarrow,-{\bf k}+{\bf q}/2}}{2}\right)^2+|\Delta|^2}\pm
\frac{\epsilon_{\uparrow,{\bf k}+{\bf q}/2}-\epsilon_{\downarrow,{\bf k}+{\bf q}/2}}{2}\\
&|u_{{\bf k},{\bf q}}|^2=\frac{1}{2}\left(1+\frac{\epsilon_{\uparrow,{\bf k}+{\bf q}/2}+\epsilon_{\downarrow,-{\bf k}+{\bf q}/2}}{E_{+,{\bf k},{\bf q}}+E_{-,{\bf k},{\bf q}}}\right)\\
&|v_{{\bf k},{\bf q}}|^2=1-|u_{{\bf k},{\bf q}}|^2,
\end{eqnarray}
and furthermore $\epsilon_{\sigma,{\bf k}}=\sum_{\alpha}2J_{\sigma,\alpha}(1-\cos(k_{\alpha}d))-\mu_{\sigma}$.
By calculating the integrals and the Matsubara summations one finds a lengthy formula
\begin{eqnarray}
\label{eq:second_pertuterm}
&-\frac{1}{2\beta \hbar^2}\int_0^{\beta\hbar}\int_0^{\beta\hbar}\,d\tau\,d\tau'\,\langle T_{\tau}\hat J(\tau)\hat J(\tau')\rangle_0\\ \nonumber
&=\sum_{\alpha,\alpha'}\frac{\Theta_{\alpha}\Theta_{\alpha'}}{M_{\alpha}M_{\alpha'}}\sum_{\bf k}\bigg[J_{\uparrow,\alpha}J_{\uparrow,\alpha'}
\sin((k_{\alpha}+q_{\alpha}/2)d)\sin((k_{\alpha'}+q_{\alpha'}/2)d)\\ \nonumber
&\times\bigg(2|u_{{\bf k},{\bf q}}|^2|v_{{\bf k},{\bf q}}|^2\frac{f(E_{+,{\bf k},{\bf q}})+f(E_{-,{\bf k},{\bf q}})-1}{E_{+,{\bf k},{\bf q}}+E_{-,{\bf k},{\bf q}}}-\\ \nonumber
&\beta|u_{{\bf k},{\bf q}}|^4 f(E_{+,{\bf k},{\bf q}})(1-f(E_{+,{\bf k},{\bf q}}))-\beta|v_{{\bf k},{\bf q}}|^4 f(E_{-,{\bf k},{\bf q}})(1-f(E_{-,{\bf k},{\bf q}}))\bigg)\\ \nonumber
&+2J_{\downarrow,\alpha}J_{\downarrow,\alpha'}\sin((k_{\alpha}-q_{\alpha}/2)d)\sin((k_{\alpha'}-q_{\alpha'}/2)d)\\ \nonumber
&\times\bigg(2|u_{{\bf k},{\bf q}}|^2|v_{{\bf k},{\bf q}}|^2\frac{f(E_{+,{\bf k},{\bf q}})+f(E_{-,{\bf k},{\bf q}})-1}{E_{+,{\bf k},{\bf q}}+E_{-,{\bf k},{\bf q}}}-\\ \nonumber 
&\beta|u_{{\bf k},{\bf q}}|^4 f(E_{-,{\bf k},{\bf q}})(1-f(E_{-,{\bf k},{\bf q}}))-\beta|v_{{\bf k},{\bf q}}|^4 f(E_{+,{\bf k},{\bf q}})(1-f(E_{+,{\bf k},{\bf q}}))\bigg)\\ \nonumber
&+4J_{\uparrow,\alpha}J_{\downarrow,\alpha'}\sin((k_{\alpha}+q_{\alpha}/2)d)\sin((k_{\alpha'}-q_{\alpha'}/2)d)\\ \nonumber
&\times|u_{{\bf k},{\bf q}}|^2|v_{{\bf k},{\bf q}}|^2\bigg(\frac{2(f(E_{+,{\bf k},{\bf q}})-f(E_{-,{\bf k},{\bf q}}))}{E_{+,{\bf k},{\bf q}}-E_{-,{\bf k},{\bf q}}}\\ \nonumber
&-\frac{2f(E_{+,{\bf k},{\bf q}})-1}{2E_{+,{\bf k},{\bf q}}}-\frac{2f(E_{-,{\bf k},{\bf q}})-1}{2E_{-,{\bf k},{\bf q}}}\bigg)\bigg], 
\end{eqnarray}
where $f(E)$ is the Fermi-Dirac distribution.
In order to handle the limit $\lim_{{\bf k'}\rightarrow {\bf k}}$, we have
implicitly assumed in equation~(\ref{eq:second_pertuterm}) that the lattice is large, i.e, $M_{\alpha}\gg 1$. 

Now the twisted grand canonical potential can be written as
\begin{equation}
\label{eq:twist_grand_pot7}
\Omega_{\bf \Theta}\approx\Omega_0+\sum_{\alpha,\alpha'}\delta\Omega_{\alpha\alpha'}\frac{\Theta_{\alpha}\Theta_{\alpha'}}{M_{\alpha}M_{\alpha'}}.
\end{equation}
In all the cases we consider $\delta\Omega_{\alpha\alpha'}=0$ when $\alpha\neq \alpha'$. The off-diagonal terms can be non-zero only when the single particle
dispersion couples the different directions together. In our case where the directions are independent, the ${\bf k}$ sums in different directions are independent, and
 because sine is an antisymmetric function these sums vanishes.
 Therefore the grand potential becomes
\begin{equation}
\label{eq:twist_grand_pot8}
\Omega_{\bf \Theta}\approx\Omega_0+\sum_{\alpha}\delta\Omega_{\alpha\alpha}\frac{\Theta_{\alpha}^2}{M_{\alpha}^2}.
\end{equation}
We can now determine the components of the dimensionless superfluid fraction tensor as
\[
\rho_{\alpha\alpha'}=\frac{\delta\Omega_{\alpha\alpha'}}{\bar{J_x}(N_{\uparrow}+N_{\downarrow})}.
\]
As a formula these components are given by
\begin{eqnarray}
\label{eq:superfluid_density}
&\rho_{\alpha\alpha}=\frac{1}{N}\sum_{{\bf k},\sigma}\tilde J_{\sigma,\alpha}\cos(k_{\alpha}d)N_{\sigma,{\bf k}}\\ \nonumber
&+\frac{1}{N}\sum_{\bf k}\bigg[\tilde J_{\uparrow,\alpha}^2\sin^2((k_{\alpha}+q_{\alpha}/2)d)\\ \nonumber
&\times\bigg(2|u_{{\bf k},{\bf q}}|^2|v_{{\bf k},{\bf q}}|^2\frac{f(E_{+,{\bf k},{\bf q}})+f(E_{-,{\bf k},{\bf q}})-1}{E_{+,{\bf k},{\bf q}}+E_{-,{\bf k},{\bf q}}}-\\ \nonumber
&\beta|u_{{\bf k},{\bf q}}|^4 f(E_{+,{\bf k},{\bf q}})(1-f(E_{+,{\bf k},{\bf q}}))-\beta|v_{{\bf k},{\bf q}}|^4 f(E_{-,{\bf k},{\bf q}})(1-f(E_{-,{\bf k},{\bf q}}))\bigg)\\ \nonumber
&+2\tilde J_{\downarrow,\alpha}^2\sin^2((k_{\alpha}-q_{\alpha}/2)d)
\bigg(2|u_{{\bf k},{\bf q}}|^2|v_{{\bf k},{\bf q}}|^2\frac{f(E_{+,{\bf k},{\bf q}})+f(E_{-,{\bf k},{\bf q}})-1}{E_{+,{\bf k},{\bf q}}+E_{-,{\bf k},{\bf q}}}-\\ \nonumber
&\beta|u_{{\bf k},{\bf q}}|^4 f(E_{-,{\bf k},{\bf q}})(1-f(E_{-,{\bf k},{\bf q}}))-\beta|v_{{\bf k},{\bf q}}|^4 f(E_{+,{\bf k},{\bf q}})(1-f(E_{+,{\bf k},{\bf q}}))\bigg)\\ \nonumber
&+4\tilde J_{\uparrow,\alpha}\tilde J_{\downarrow,\alpha'}\sin((k_{\alpha}+q_{\alpha}/2)d)\sin((k_{\alpha}-q_{\alpha}/2)d)\\ \nonumber
&\times|u_{{\bf k},{\bf q}}|^2|v_{{\bf k},{\bf q}}|^2\bigg(\frac{2(f(E_{+,{\bf k},{\bf q}})-f(E_{-,{\bf k},{\bf q}}))}{E_{+,{\bf k},{\bf q}}-E_{-,{\bf k},{\bf q}}}\\ \nonumber
&-\frac{2f(E_{+,{\bf k},{\bf q}})-1}{2E_{+,{\bf k},{\bf q}}}-\frac{2f(E_{-,{\bf k},{\bf q}})-1}{2E_{-,{\bf k},{\bf q}}}\bigg)\bigg]\\ \nonumber
&\rho_{\alpha\alpha'}=0,\nonumber
\end{eqnarray}
where $N=N_{\uparrow}+N_{\downarrow}$ and $\tilde J_{\sigma,\alpha}=J_{\sigma,\alpha}/\bar{J_x}$.
When the momentum ${\bf q}$ is non-zero the superfluid density describes the one mode FFLO phase, and
when the momentum ${\bf q}=0$ the superfluid density describes the BCS or the Sarma phase.
$\rho_{\alpha\alpha}$ is a superfluid fraction, it describes what fraction of the atoms is in the superfluid state.
In the limit where $J_{\uparrow,\alpha}=J_{\downarrow,\alpha'}=J$ for all $\alpha,\alpha'$  and $N_{\uparrow}=N_{\downarrow}$,
the chemical potentials are same and ${\bf q}=0$.
In this limit superfluid density tensor can be simplified as
\begin{eqnarray}
\label{eq:superfluid_density_2}
&\rho_{\alpha\alpha}=\frac{1}{N}\sum_{{\bf k},\sigma}\cos(k_{\alpha}d)N_{\sigma,{\bf k}}-\frac{4\beta J}{N}\sum_{\bf k}\sin^2(k_{\alpha}d)f(E_{\bf k})(1-f(E_{\bf k}))\\
&\rho_{\alpha\alpha'}=0
\end{eqnarray}
where 
\[E_{\bf k}=\sqrt{\left(\frac{\epsilon_{\uparrow,{\bf k}}+\epsilon_{\downarrow,-{\bf k}}}{2}\right)^2+|\Delta|^2}.\]

In the continuum limit i.e. $d\rightarrow 0$, where $Jd^2$ remains constant, the superfluid fraction becomes
\[\rho_{\alpha\alpha}=\rho=1+\frac{2\hbar^2}{m_anV}\sum_{\bf k}k_z^2\frac{\partial f(E_{\bf k})}{\partial E_{\bf k}},\]
where the effective mass $m_a=\hbar^2/(2Jd^2)$, $n=n_{\uparrow}+n_{\downarrow}$ is the total number density and $V$ is the size of the system.
This continuum result is precisely the Landau's formula for the BCS superfluid fraction~\cite{Lifshitz2002a}.

\section{Results}
\label{sec:results}

\subsection{BCS/Sarma-phase results}

\begin{figure}[!htb]
\centering
\begin{minipage}{0.45\textwidth}
\centering 
\includegraphics[width=\textwidth]{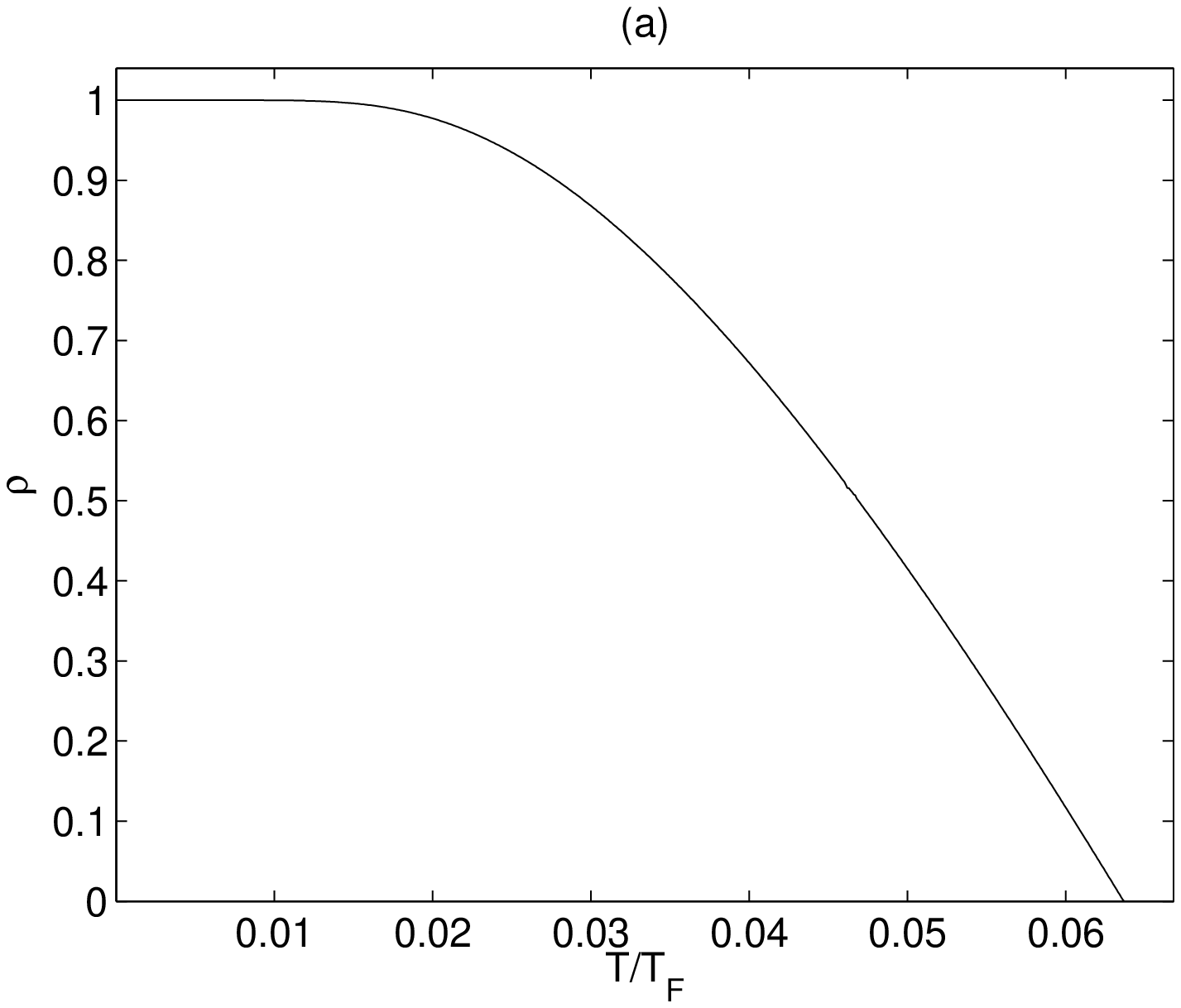}
\end{minipage}
\begin{minipage}{0.45\textwidth}
\centering 
\includegraphics[width=\textwidth]{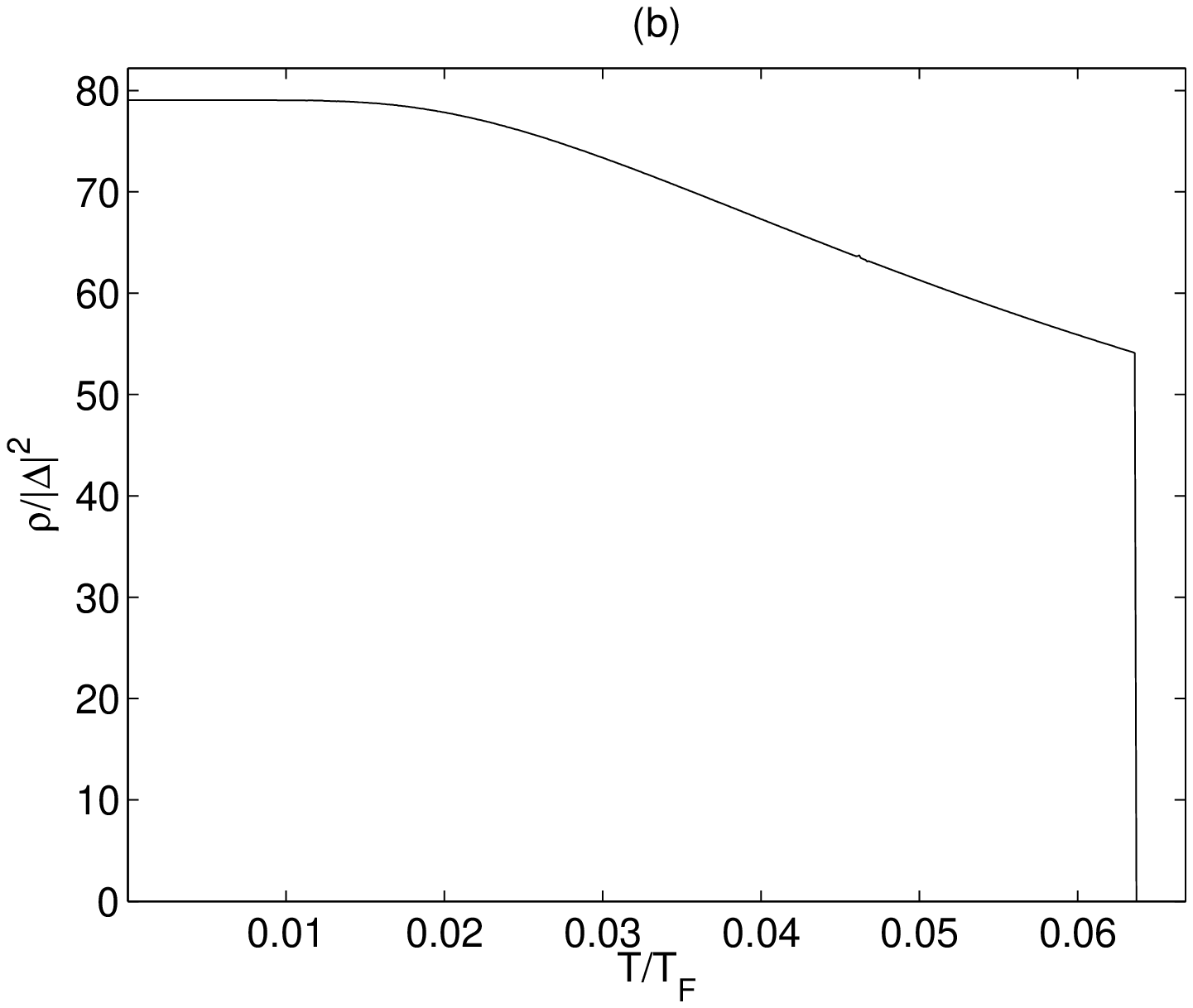}
\end{minipage}
\caption{
Figure (a)  we show the BCS superfluid fraction as a function of the temperature in
a free space (absence of the lattice).
Figure (b) shows  the BCS superfluid fraction divided by $|\Delta|^2$ as a function
of temperature. The interaction strength $k_Fa_s=-0.30$, where $k_F$ is the
Fermi wave vector. 
}
\label{fig:fig1}
\end{figure}

In the continuum the formula of the BCS superfluid fraction is the well known Landau's formula.
Figure~\ref{fig:fig1} (a) shows the BCS superfluid fraction as a function of the temperature, in the uniform case (the external potential is zero), without
the lattice. In figure~\ref{fig:fig1} (b) we show the BCS superfluid fraction divided by $|\Delta|^2$ as a a function of the temperature.
One sees from figure~\ref{fig:fig1} (a) that the BCS superfluid fraction is one at zero temperature and from figure~\ref{fig:fig1} (b) that
the superfluid fraction is almost proportional to $|\Delta|^2$, but the temperature dependence of the superfluid fraction differs somewhat from the temperature dependence of
 $|\Delta|^2$. The standard BCS result is that the superfluid fraction is proportional to $|\Delta|^2$ in limit $T\rightarrow T_c$~\cite{FetterWalecka}.
When we calculated the gap, we used the renormalization, i.e., we have cancelled out the divergent part of the gap equation,
in reference~\cite{FetterWalecka} is used the cutoff energy. This is the cause of little difference between the results.

When all the hopping strengths are the same i.e. $J=J_{\uparrow,\alpha}=J_{\downarrow,\alpha'}$, and  filling fractions $n_{\sigma}=N_{\sigma}/M_xM_yM_z$ are same
and the average filling fraction $n_{av}=(n_{\uparrow}+n_{\downarrow})/2$ goes to zero, the superfluid fraction should go to one at zero temperature.
For low filling fractions the the atoms occupy only the lowest momentum states and 
the single particle dispersions  $\epsilon_{\sigma,{\bf k}}=\sum_{\alpha}2J_{\sigma,\alpha}(1-\cos(k_{\alpha}d))-\mu_{\sigma}$
approaches to form $J(kd)^2-\mu_{\sigma}$, which is the  free space dispersion. Therefore the results on the limit of low filling fractions become
the free space result and  we know from the Landau's formula that the BCS superfluid fraction in the free
space is one at zero temperature (see figure~\ref{fig:fig1} (a)) . 

\begin{figure}[!htb]
\centering
\begin{minipage}{0.45\textwidth}
\centering 
\includegraphics[width=\textwidth]{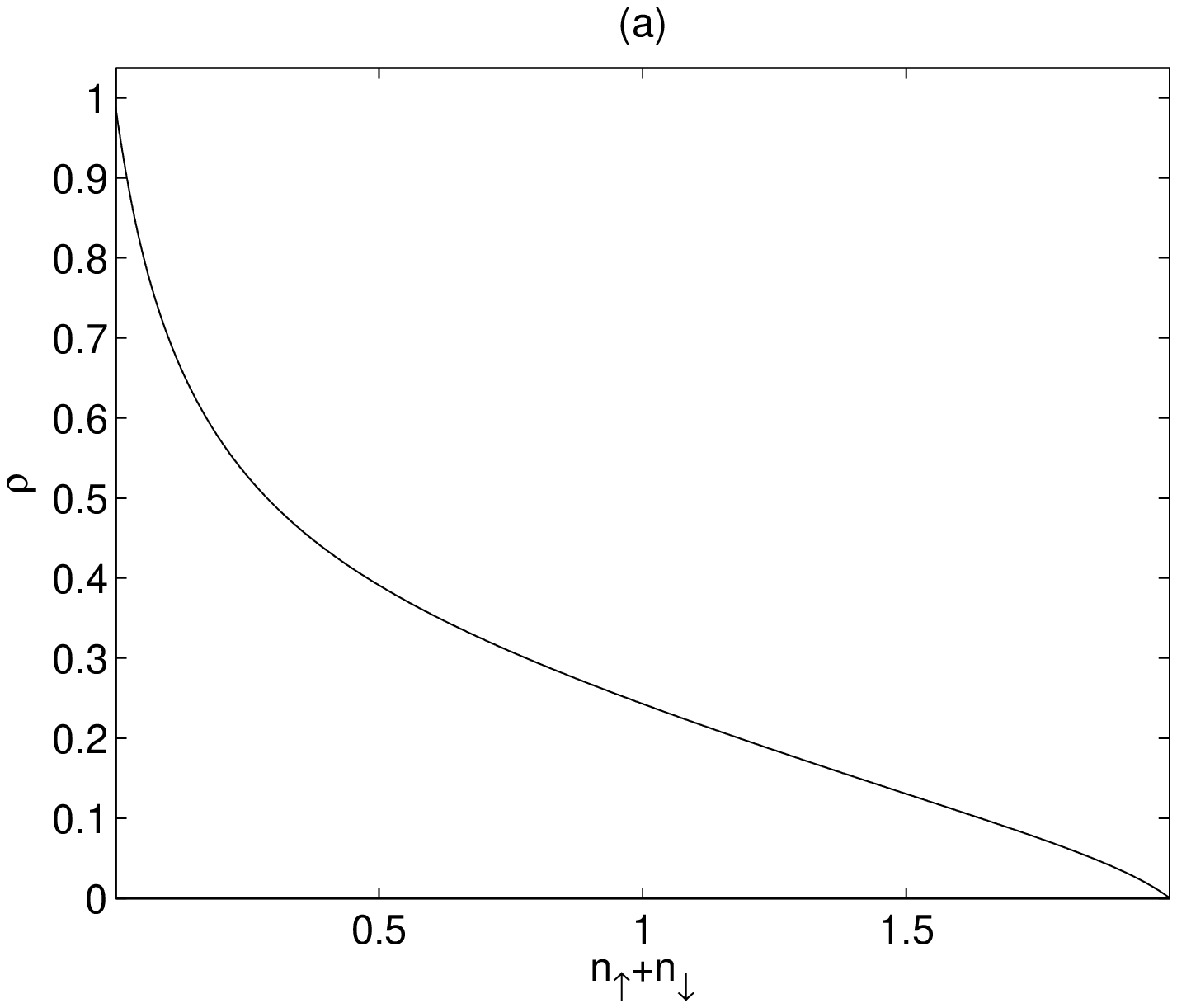}
\end{minipage}
\begin{minipage}{0.45\textwidth}
\centering 
\includegraphics[width=\textwidth]{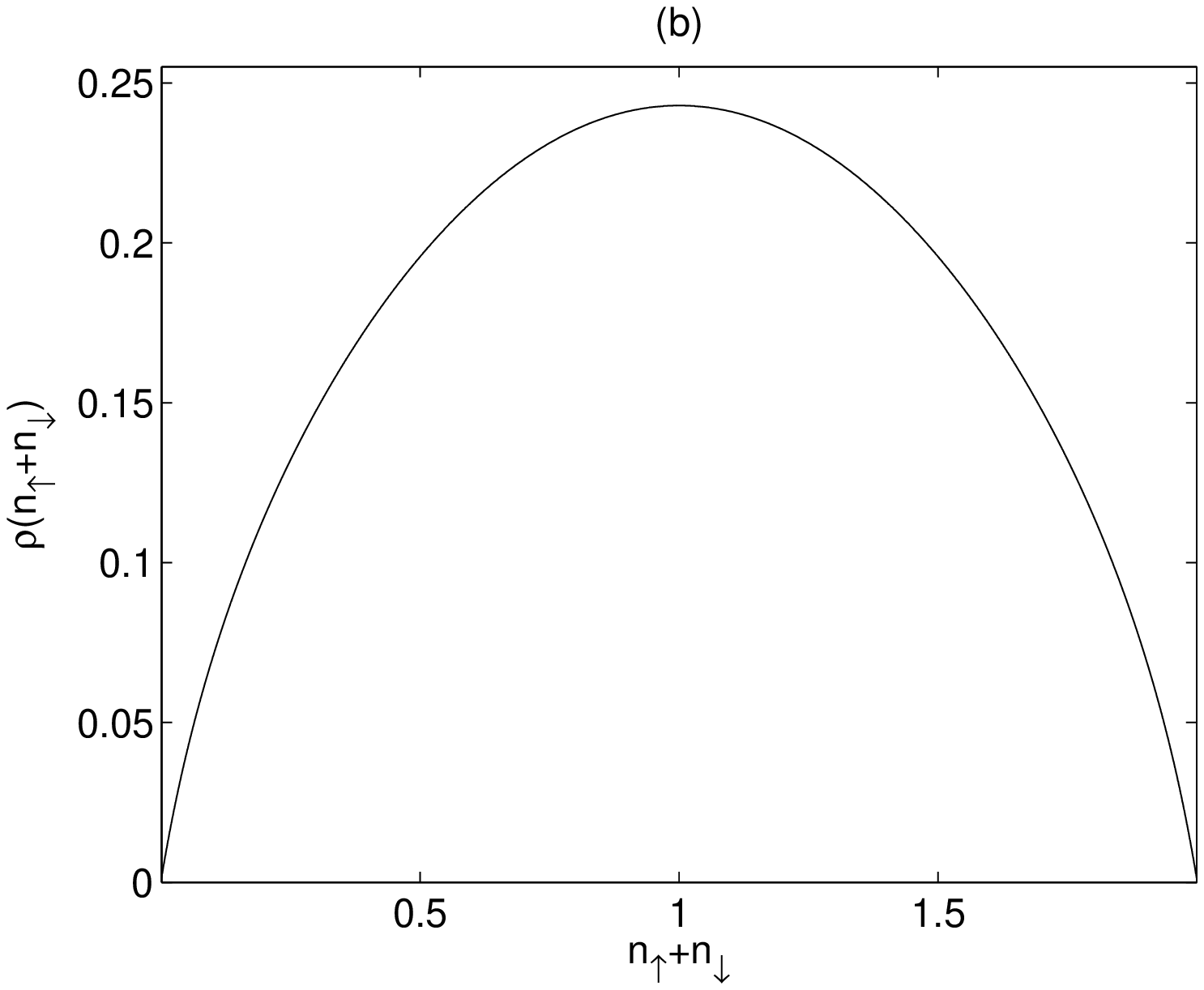}
\end{minipage}
\begin{minipage}{0.45\textwidth}
\centering 
\includegraphics[width=\textwidth]{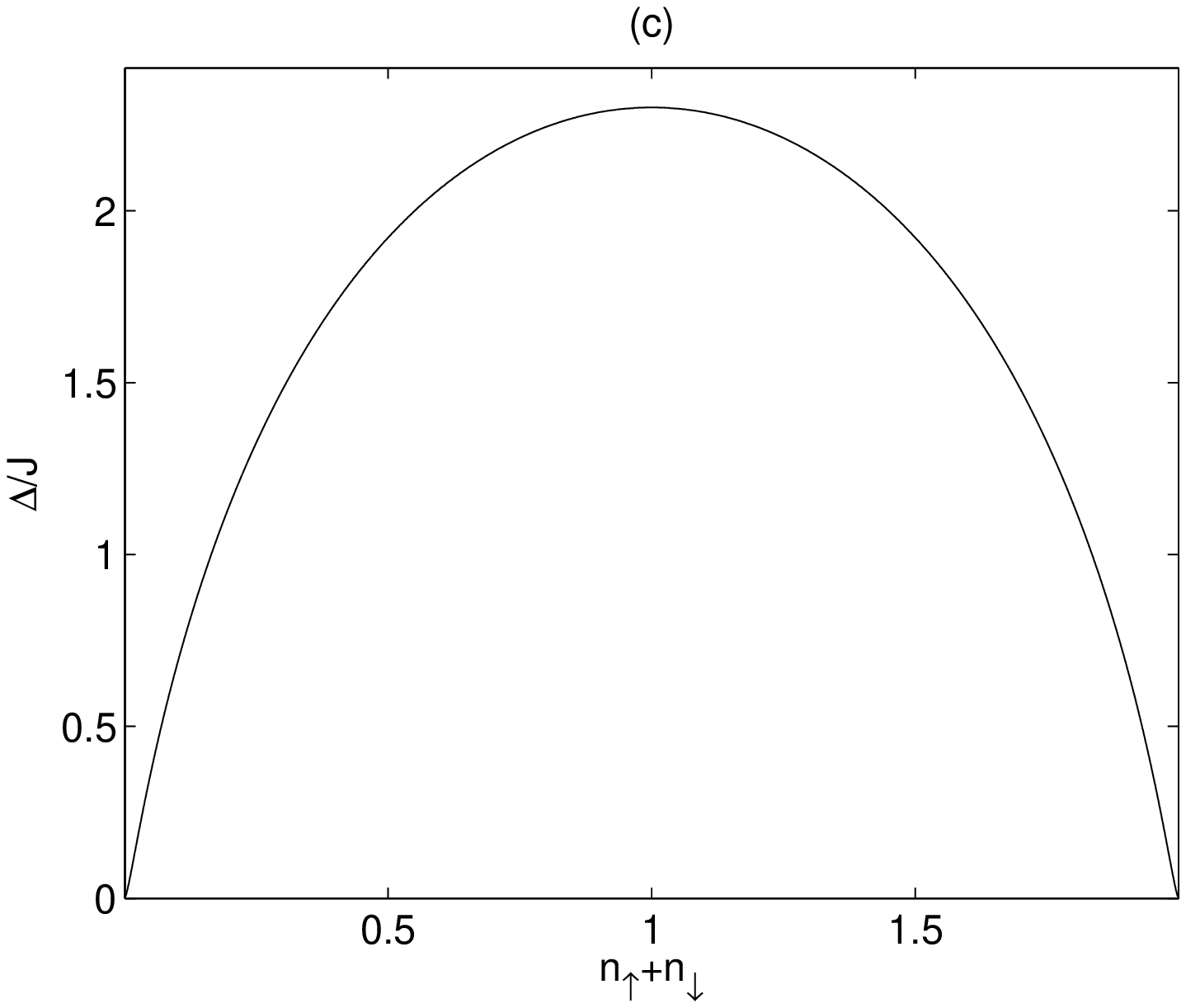}
\end{minipage}
\begin{minipage}{0.45\textwidth}
\centering 
\includegraphics[width=\textwidth]{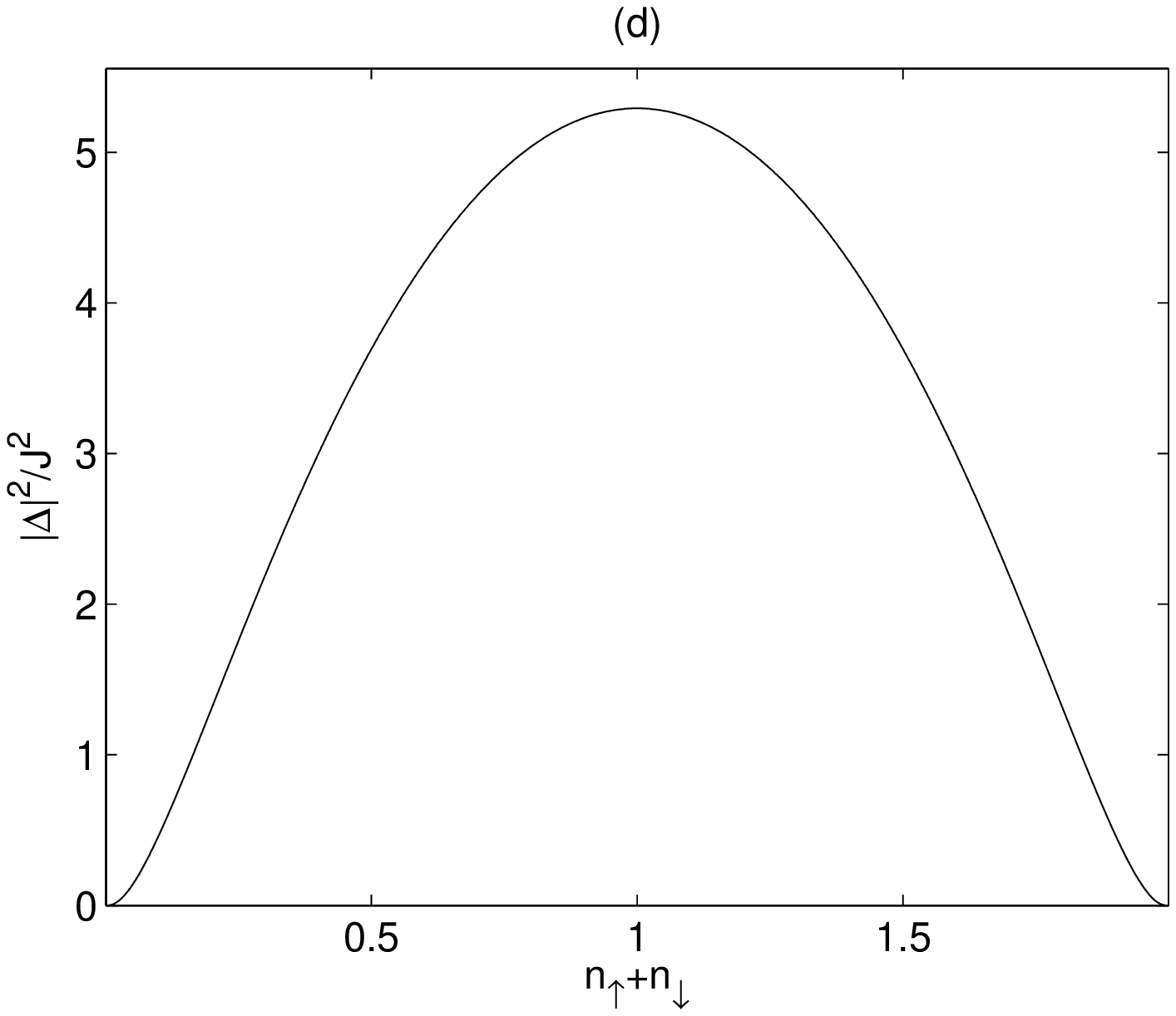}
\end{minipage}
\caption{
In figure (a) we show the BCS superfluid fraction $\rho=\rho_{xx}=\rho_{yy}=\rho_{zz}$ as a function of the total
filling fraction $n_{\uparrow}+n_{\downarrow}=2n_{\uparrow}=2n_{\downarrow}$ at zero temperature. In figure $(b)$ is shown
$\rho(n_{\uparrow}+n_{\downarrow})$ as a function of the total filling fraction. Figures (c) and (d) show $\Delta/J$ and $|\Delta|^2/J^2$, respectively as functions
of the total filling fraction.
All the hopping strengths are same i.e. $J=J_{\uparrow,\alpha}=J_{\downarrow,\alpha'}$,
 and $-U/J=6.0$.}
\label{fig:fig2}
\end{figure}

In figure~\ref{fig:fig2} (a) we show the BCS superfluid fraction as a function of the total
filling fraction in the lattice. As we can see, the BCS superfluid fraction  approaches one, when the filling fractions become small.
In the free space the superfluid fraction is one at zero temperature, but in the lattice
this no longer holds.
In figure~\ref{fig:fig2} (b) we show $\rho(n_{\uparrow}+n_{\downarrow})$  as a function of the total  filling fraction and as
one can see from it, the BCS superfluid density, which is not divided by the total filling fraction is symmetric with respect to half filling.
The cause of this is the particle-hole symmetry of the lattice model.
The particle-hole symmetry of the lattice model can be seen also from figures~\ref{fig:fig2} (c) and~\ref{fig:fig2} (d). In
figures~\ref{fig:fig2} (c) and~\ref{fig:fig2} (d) we show $\Delta/J$ and $|\Delta|^2/J^2$, respectively, as functions of 
of the total filling fraction.
By comparing figures~\ref{fig:fig2} (b)-(d) one can notice that $\rho(n_{\uparrow}+n_{\downarrow})$ is not directly proportional to $\Delta$ or $|\Delta|^2$.

\begin{figure}[!htb]
\centering
\begin{minipage}{0.45\textwidth}
\centering 
\includegraphics[width=\textwidth]{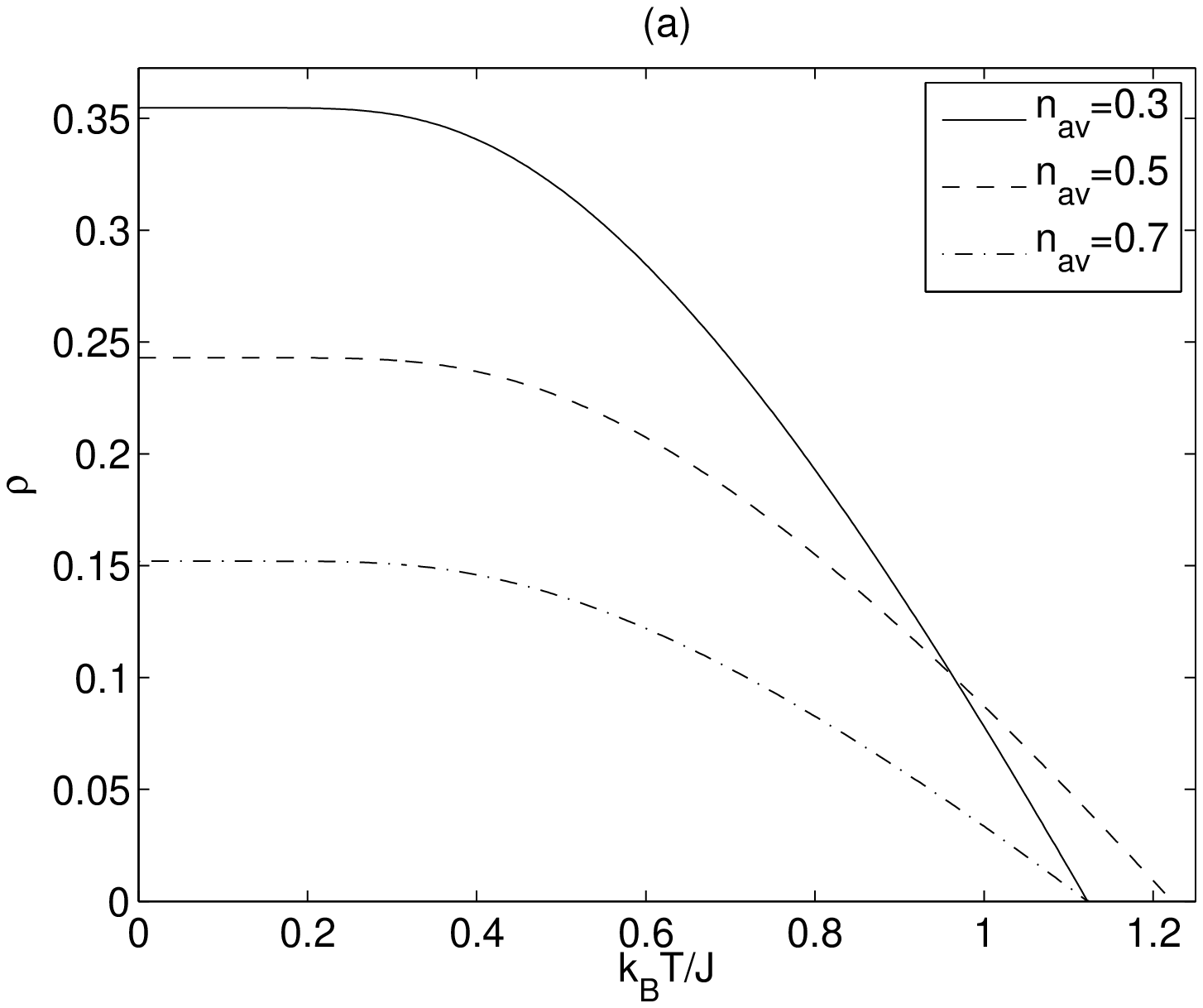}
\end{minipage}
\begin{minipage}{0.45\textwidth}
\centering 
\includegraphics[width=\textwidth]{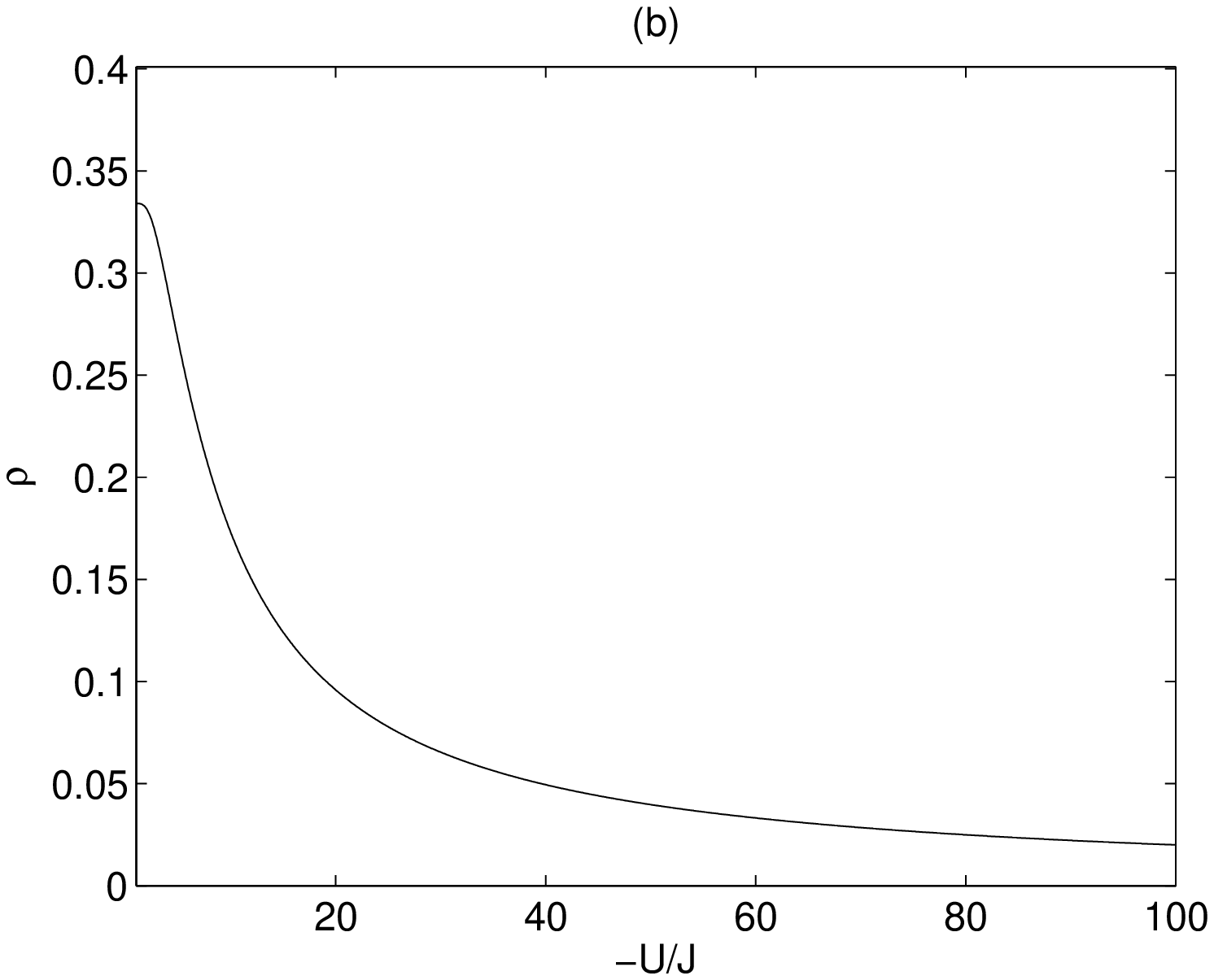}
\end{minipage}
\begin{minipage}{0.45\textwidth}
\centering 
\includegraphics[width=\textwidth]{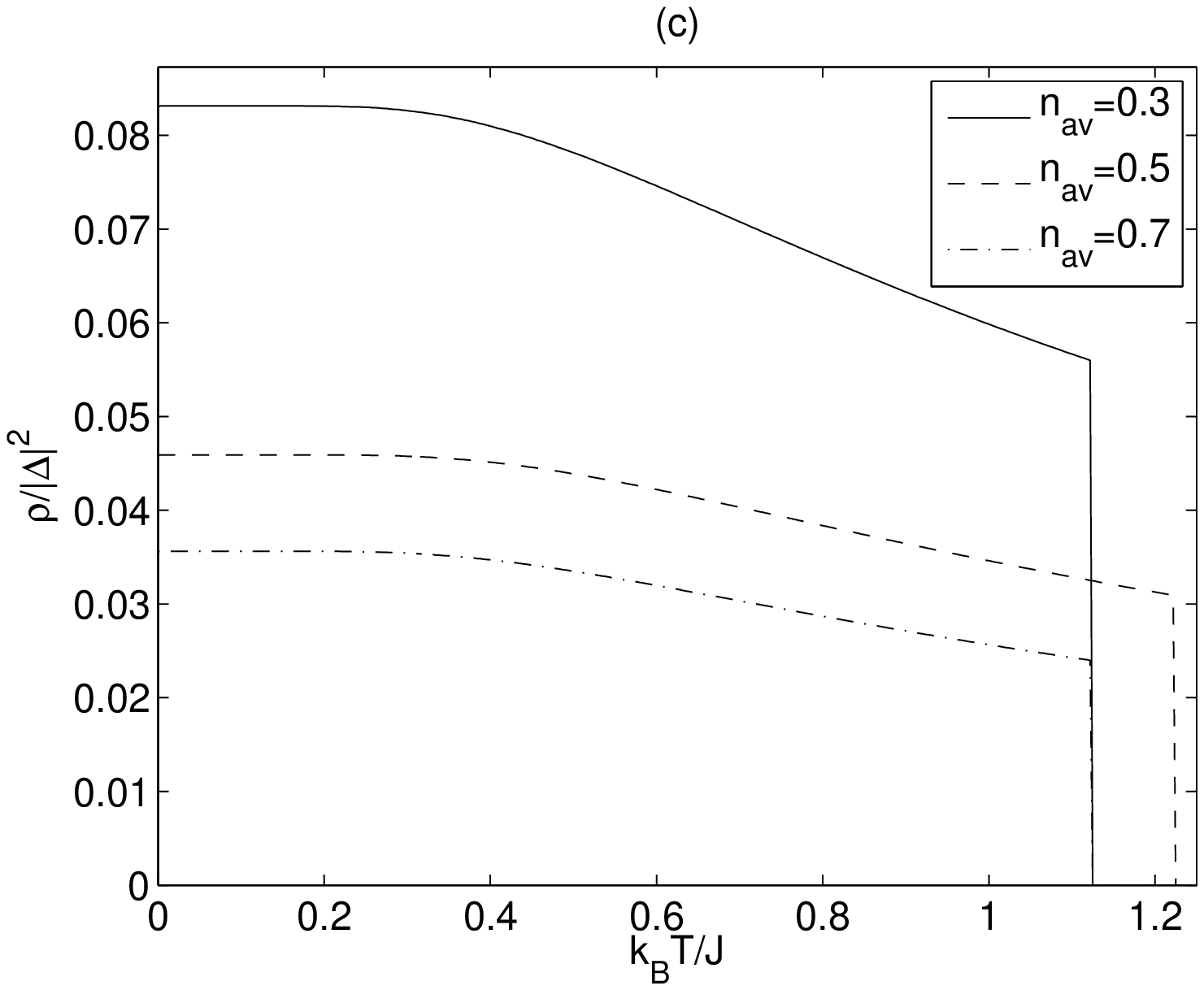}
\end{minipage}
\begin{minipage}{0.45\textwidth}
\centering 
\includegraphics[width=\textwidth]{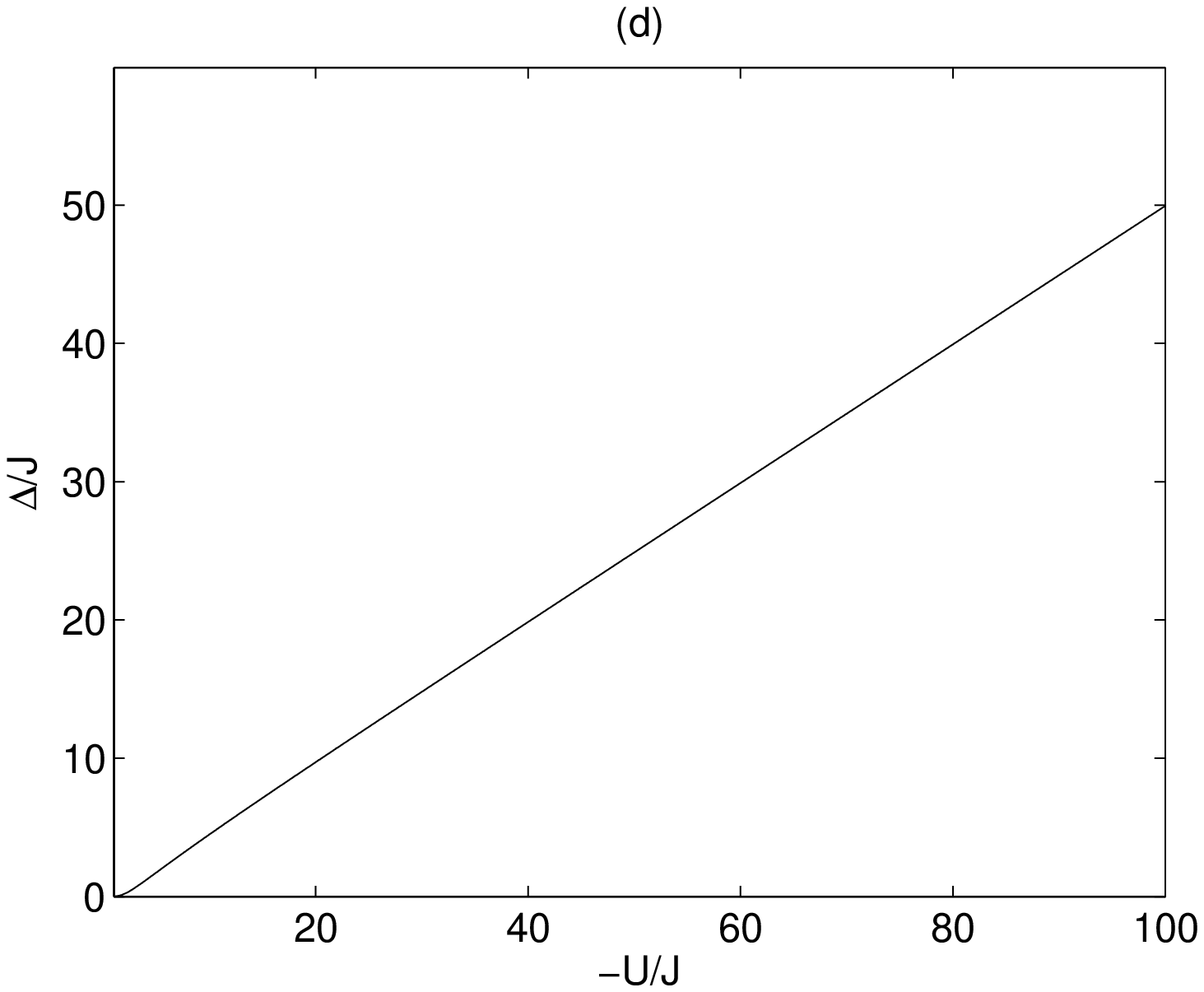}
\end{minipage}
\caption{
Figure (a) shows  the BCS superfluid fraction $\rho=\rho_{xx}=\rho_{yy}=\rho_{zz}$ as a function of the temperature,with three
different average filling fractions $n_{av}=(n_{\uparrow}+n_{\downarrow})/2$. Figure (b) shows  the BCS superfluid fraction as a function of a  dimensionless
coupling strength  $-U/J$ at zero temperature.
In figure (c) we show the BCS superfluid fraction divided by $|\Delta|^2$ as a function of the temperature,with three
different average filling fractions.
Figure (d)  shows the pairing gap as a function of $-U/J$ at zero temperature. 
In the all figures all the hopping strengths are same i.e. $J=J_{\uparrow,\alpha}=J_{\downarrow,\alpha'}$.
In figures (a) and (c) $-U/J=6.0$, and in figures (b) and (d) $n_{\uparrow}=n_{\downarrow}=0.5$.}
\label{fig:fig3}
\end{figure}

Figure~\ref{fig:fig3} (a) shows the BCS superfluid fraction as a function of the temperature, with three
different average filling fractions. When one compares this figure to figure~\ref{fig:fig1} (a), one notices
that the lattice result is very similar compared to the free space result. The superfluid fraction goes to zero, when the paring gap goes to zero as expected. 
The figure is consistent with figure~\ref{fig:fig2} (a), in that
the total filling fraction increases when the superfluid fraction decreases.
In figure~\ref{fig:fig3} (b) we show the BCS superfluid fraction as a function of the dimensionless coupling strength  $-U/J$ at zero temperature.
As we can see, when $-U/J$ increases the superfluid fraction decreases contrary to the free space result where superfluid fraction
is a constant as a function of the coupling strength at zero temperature. Of course at finite temperatures
the free space superfluid fraction decreases with increasing coupling strength.
This is due to the fact as $-U/J$ increases the movement of the atoms decreases in the lattice. 
In other words when $-U/J$ increases then Cooper pairs become smaller and the atoms are better localized.
Of course in the limit $-U/J\rightarrow\infty$ the mean-field theory fails and in large values of $-U/J$ the results are not reliable.
Figure~\ref{fig:fig3} (c) shows the BCS superfluid fraction divided by $|\Delta|^2$ as a function of the temperature, with three
different average filling fractions and we see that the BCS superfluid fraction is almost proportional to $|\Delta|^2$ (compare to figure~\ref{fig:fig1} (b)) i.e 
$\rho=c(T,n_{\uparrow}=n_{\downarrow})|\Delta|^2$,
where $c(T,n_{\uparrow}=n_{\downarrow})$ depends weakly on the temperature.
We show the pairing gap as a function of $-U/J$ at zero temperature in figure~\ref{fig:fig3} (d). As one can notice, the BCS superfluid fraction does not
behave anything like the pairing gap as a function of $-U/J$ at zero temperature.

\begin{figure}[!htb]
\centering
\begin{minipage}{0.45\textwidth}
\centering 
\includegraphics[width=\textwidth]{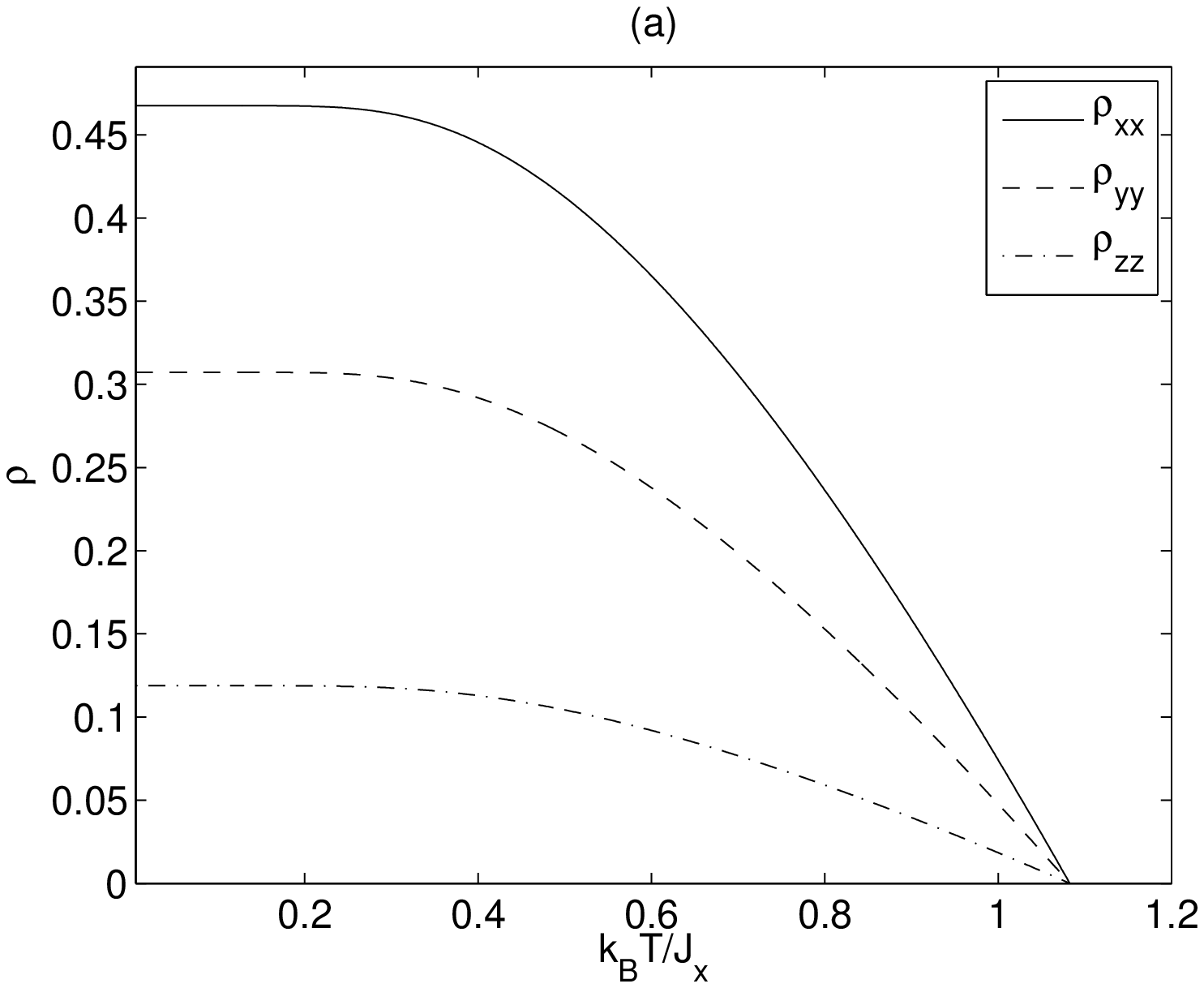}
\end{minipage}
\begin{minipage}{0.45\textwidth}
\centering 
\includegraphics[width=\textwidth]{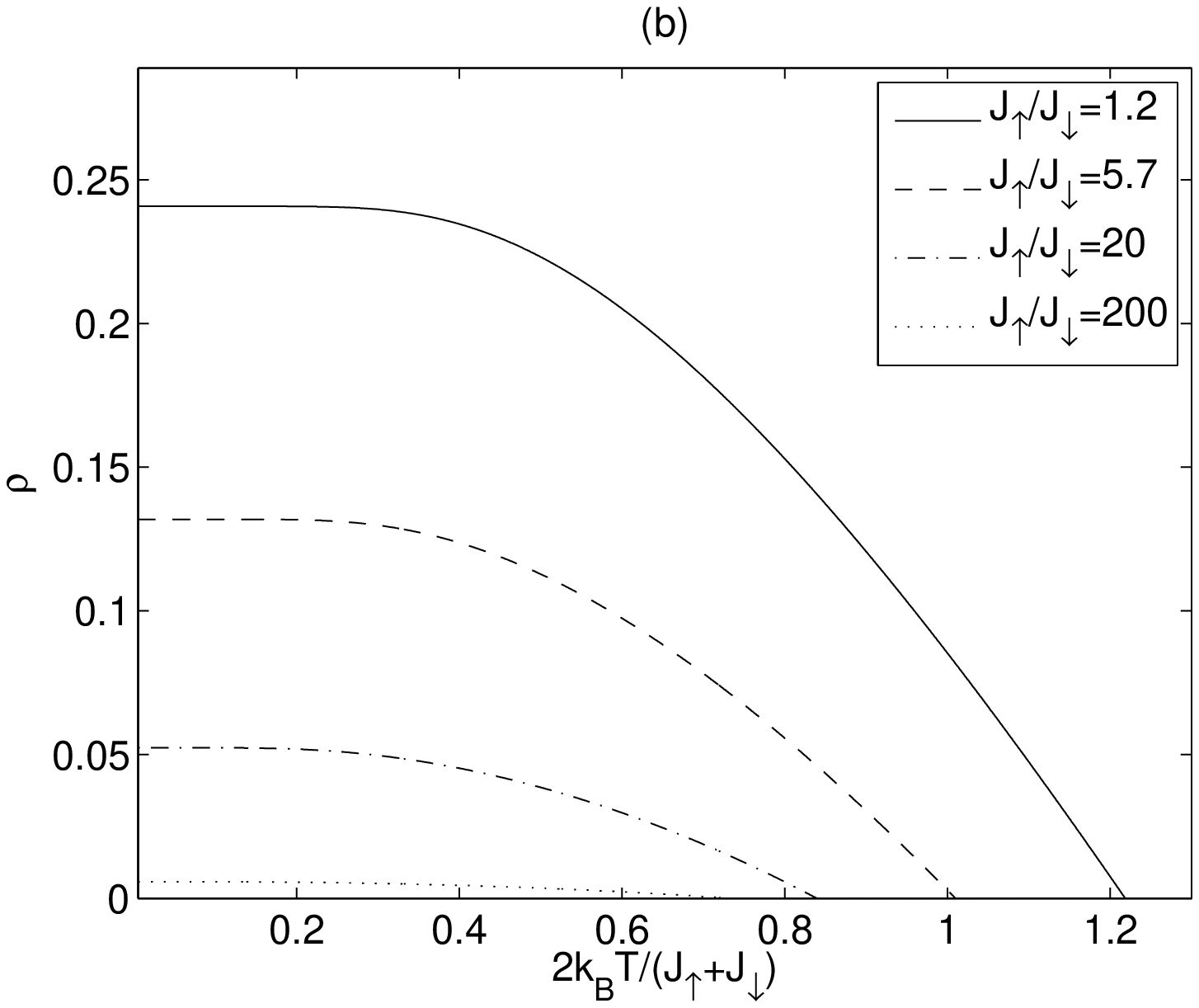}
\end{minipage}
\begin{minipage}{0.45\textwidth}
\centering 
\includegraphics[width=\textwidth]{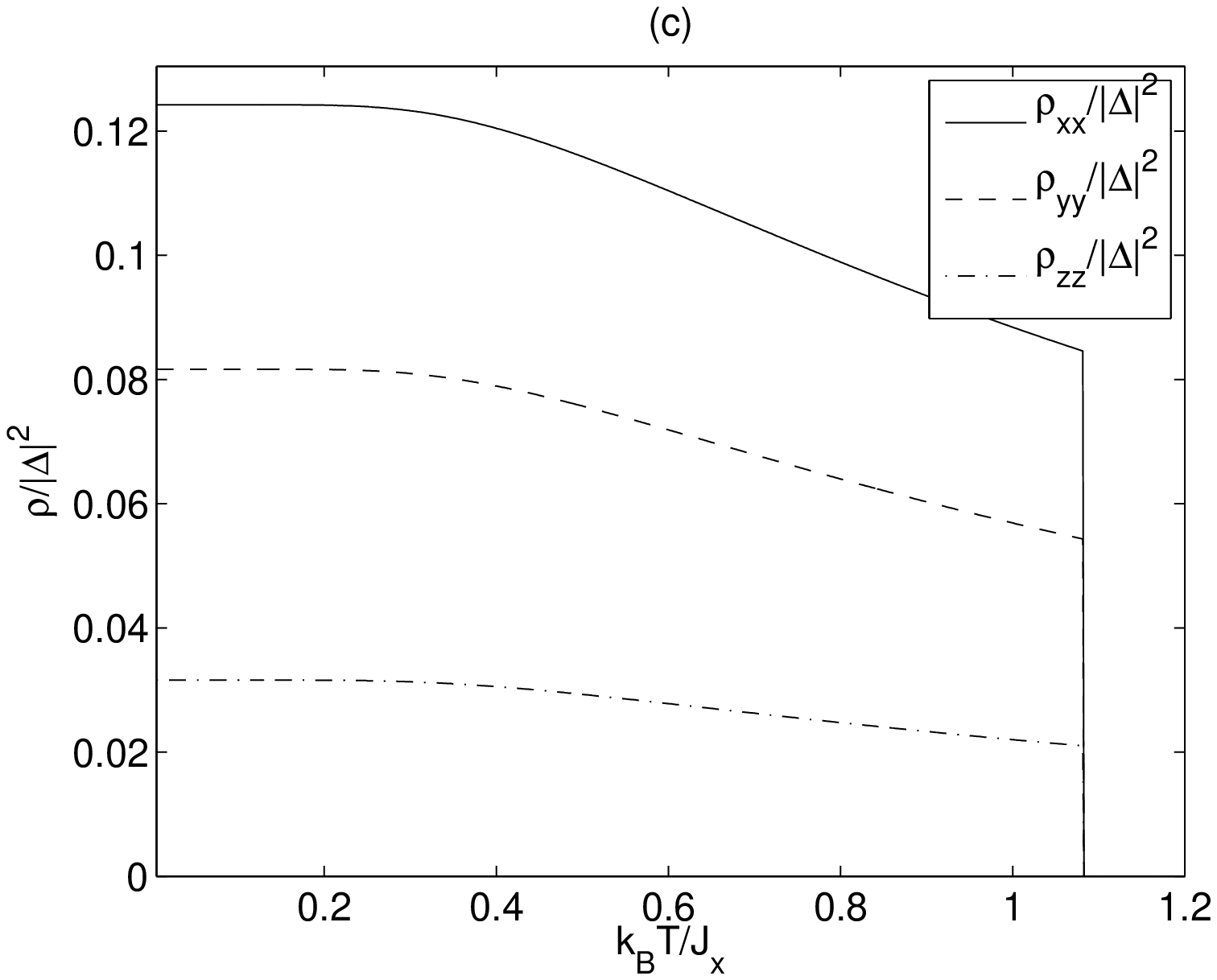}
\end{minipage}
\begin{minipage}{0.45\textwidth}
\centering 
\includegraphics[width=\textwidth]{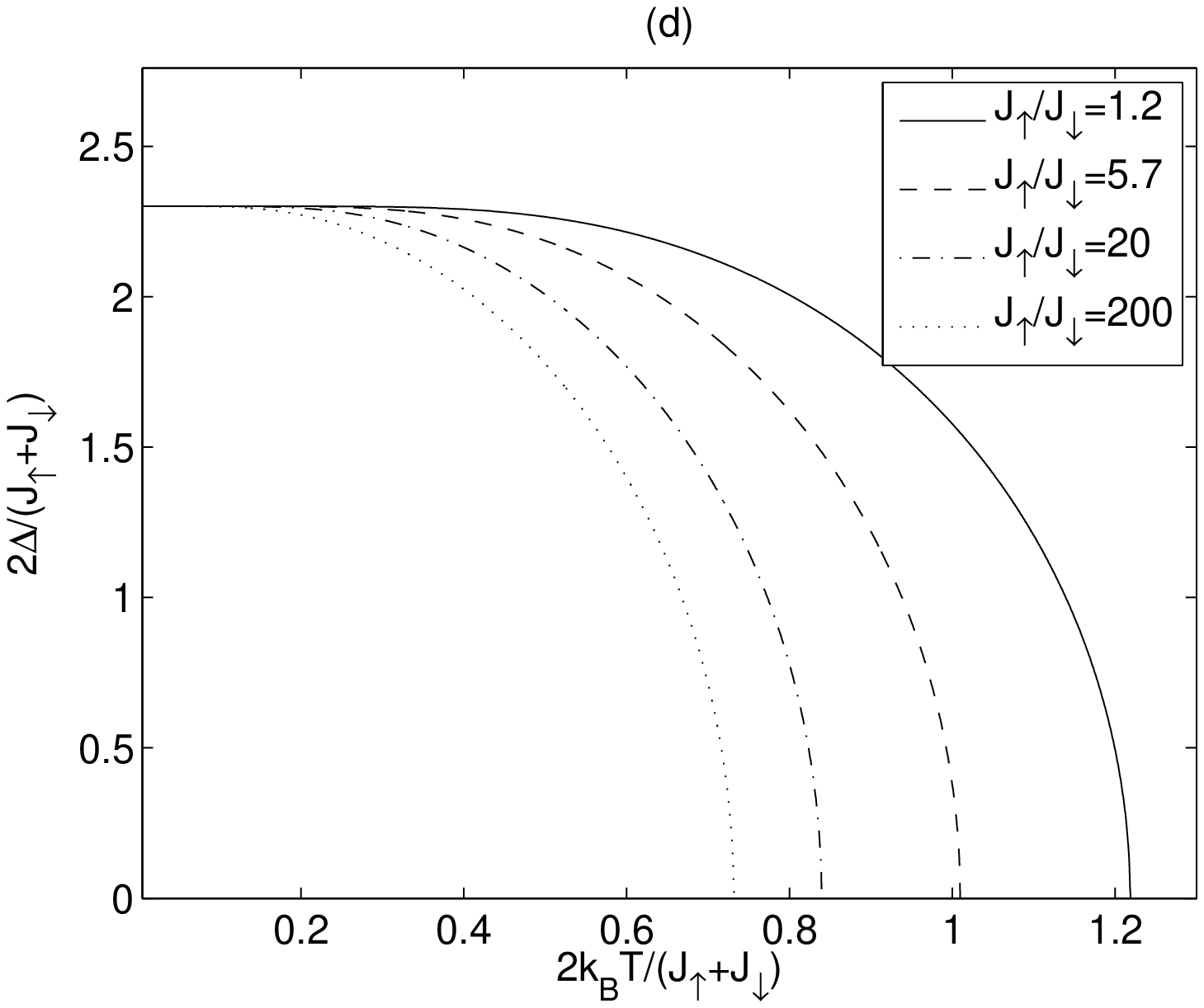}
\end{minipage}
\caption{[Colour online]
In figure (a) we show the BCS superfluid fraction as a function of the temperature in the case, where the hopping strengths are different in
the different directions.
In figure (b) we  show the BCS superfluid fraction as a function of the temperature.
Figure (c) shows the BCS superfluid fraction divided by $|\Delta|^2$ as a function of 
the temperature.
In figure (d)  we show the pairing gaps as functions of the temperature.
In figures (a) and (c) $J_x=(J_{\uparrow,x}+J_{\downarrow,x})/2$,  
$J_{\uparrow,y}=J_{\downarrow,y}=0.8J_{\uparrow,x}=0.8J_{\downarrow,x}$, $J_{\uparrow,z}=J_{\downarrow,z}=0.5J_{\uparrow,x}=0.5J_{\downarrow,x}$, 
$n_{\uparrow}=n_{\downarrow}=0.2$, and $-U/J_{\uparrow,x}=6.0$.
In figures (b) and (d) $J_{\sigma,\alpha}=J_{\sigma',\alpha'}=J_{\sigma}$,  $-2U/(J_{\uparrow}+J_{\downarrow})=6.0$, and  $n_{\uparrow}=n_{\downarrow}=0.5$}
\label{fig:fig4}
\end{figure}

Since it is experimentally easy to study anisotropic lattices and use these to explore dimensional crossovers, let us explore superfluid fractions in 
anisotropic lattices.
We show in figure~\ref{fig:fig4} (a) the BCS superfluid fraction as a function of the temperature in the case, where the hopping strengths are different in
the different directions. As one can see from the figure the components of the superfluid fraction tensor are different in different directions.
Furthermore, we see that the superfluid fraction is larger in the direction of large hopping strength. This is easy to understand, and is
due to the free energy difference in $\alpha$-direction being proportional to $J_{\alpha}$ as we see from equations~(\ref{eq:twist_grand_pot8})
and~(\ref{eq:superfluid_density}). If the superfluid fraction was defined as in equation~(\ref{eq:2_maar}), the components of the
superfluid fraction would be equal in every direction.
In other words the effective masses are different in different directions.
figure~\ref{fig:fig4} (c) shows that in the in the case, where the hopping strengths are different in the different directions, $\rho_{\alpha\alpha}/|\Delta|^2$ is almost 
a constant as a function of the temperature.

Figure~\ref{fig:fig4} (b) shows the BCS superfluid fraction as a function of the temperature in the case, where the hopping strengths are different for the components
(but $J_{\uparrow}+J_{\downarrow}$ is kept constant).
As it is clear from  figure~\ref{fig:fig4} (b), when the ratio $J_{\uparrow}/J_{\downarrow}$ increases the superfluid fraction decreases, but the pairing gap, however,
does not decrease as a function of $J_{\uparrow}/J_{\downarrow}$ at zero temperature (it does as a function of the temperature), 
but it remains almost a constant  (at zero temperature it is a constant). This can be seen from figure~\ref{fig:fig4} (d), in which is shown
the pairing gaps as as functions of the temperature in the cases, where the hopping strengths are different for the components.
 One can also notice that when $J_{\uparrow}/J_{\downarrow}\rightarrow \infty$ then $\rho \rightarrow 0$ ( in this case the critical
temperature also goes to zero).
When $J_{\uparrow}/J_{\downarrow}$ increases but $2U/(J_{\uparrow}+J_{\downarrow})$ remains constant, the lattice becomes deeper and deeper
for $\downarrow$-component. Thus the atoms of the component $\downarrow$ are more localized and the atoms do not move easily. 

\begin{figure}[!htb]
\centering
\begin{minipage}{0.45\textwidth}
\centering 
\includegraphics[width=\textwidth]{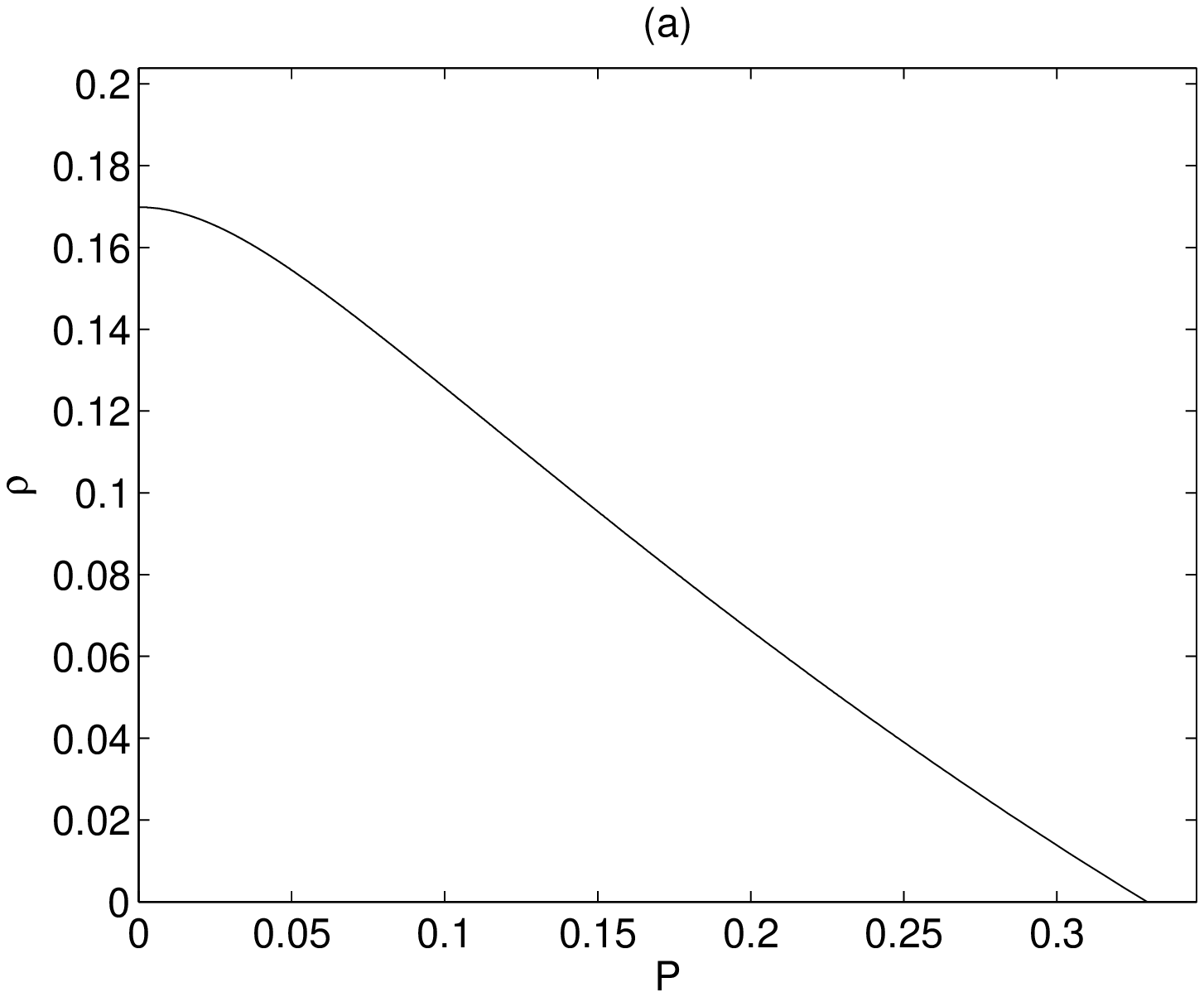}
\end{minipage}
\begin{minipage}{0.45\textwidth}
\centering 
\includegraphics[width=\textwidth]{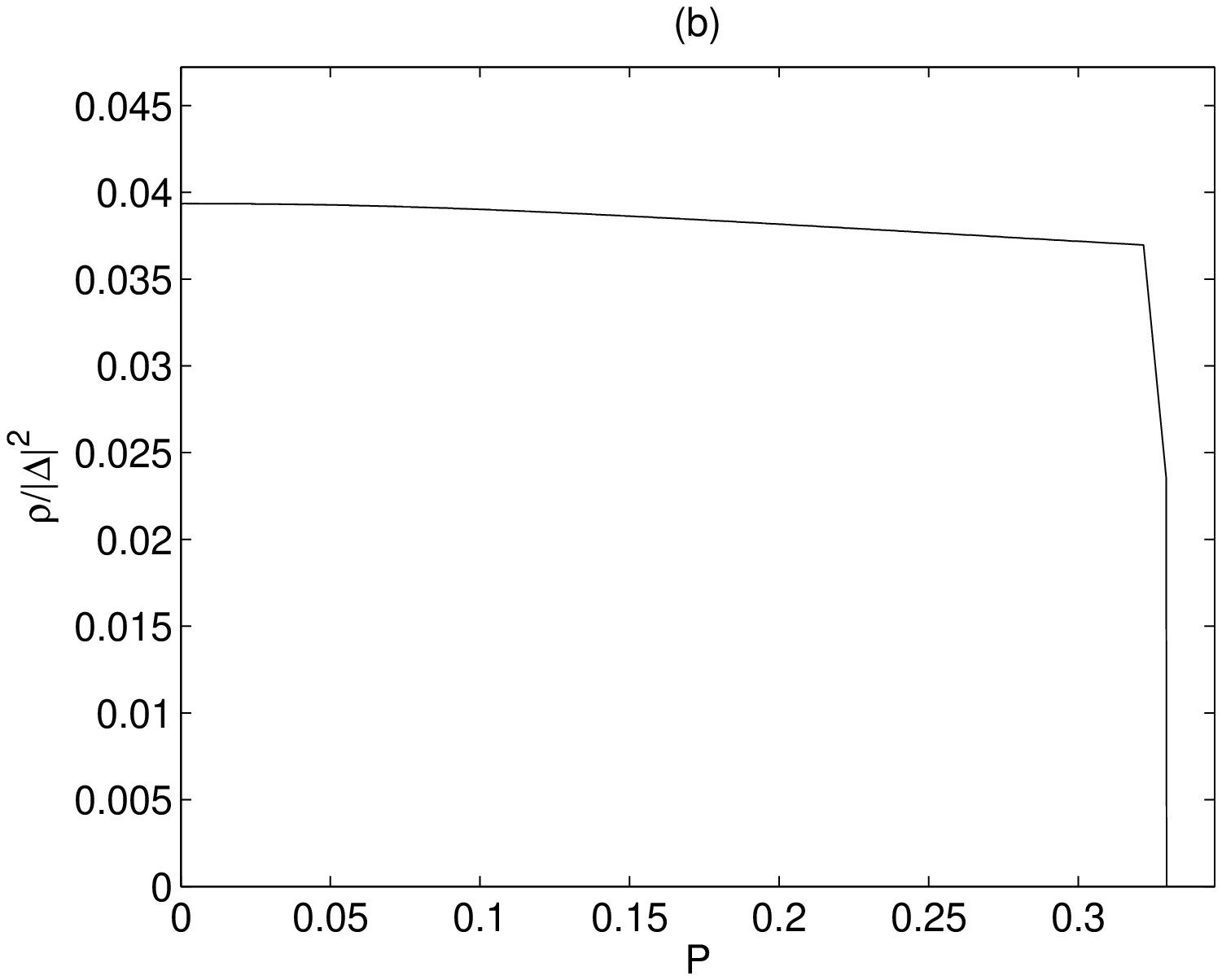}
\end{minipage}
\begin{minipage}{0.45\textwidth}
\centering 
\includegraphics[width=\textwidth]{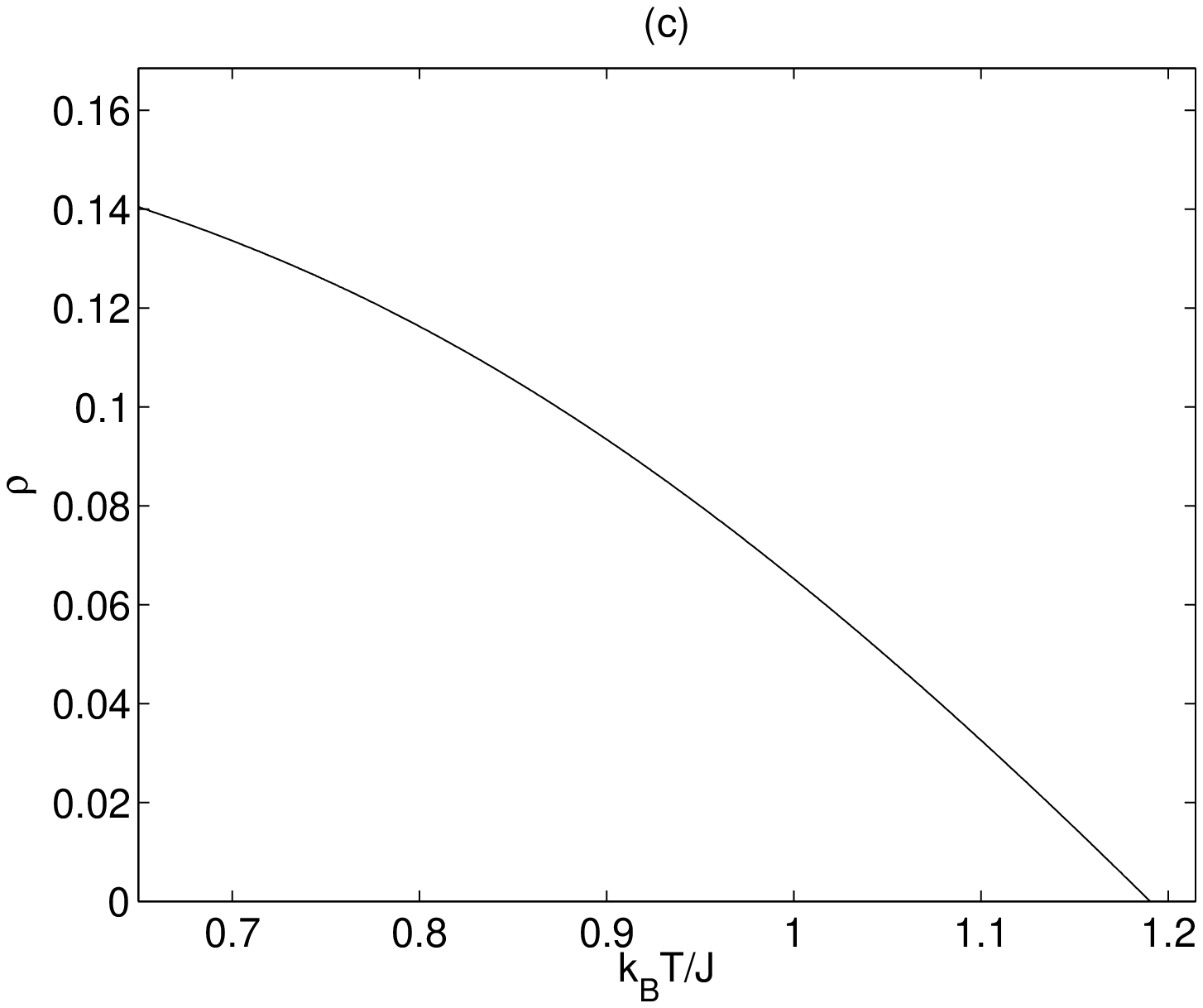}
\end{minipage}
\begin{minipage}{0.45\textwidth}
\centering 
\includegraphics[width=\textwidth]{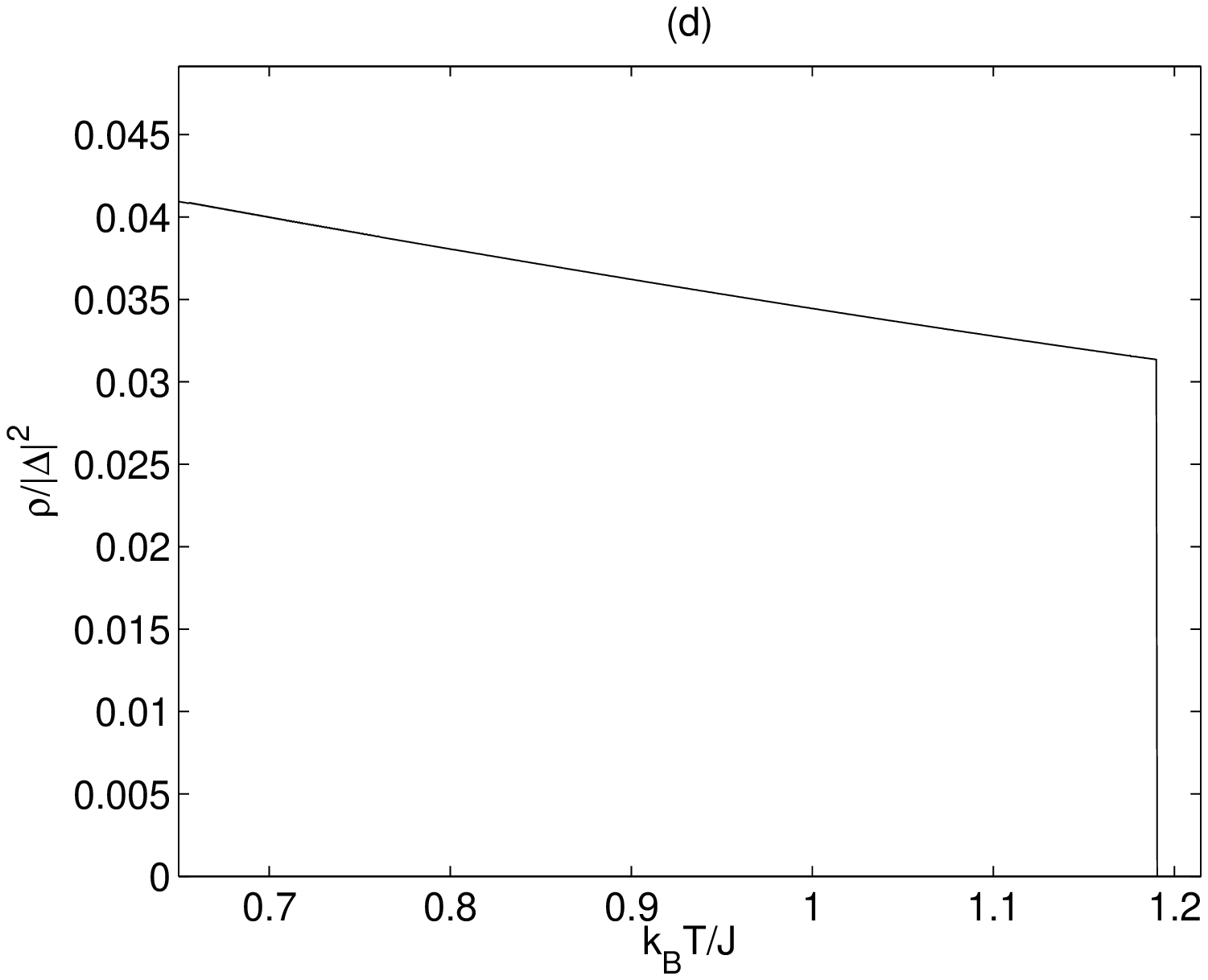}
\end{minipage}
\caption{[Colour online] The BP/Sarma phase superfluid fraction.
In figure (a) we show the superfluid fraction as a function of the polarization $P=(n_{\uparrow}-n_{\downarrow})/(n_{\uparrow}-n_{\downarrow})$.
Figure (b) shows the superfluid fraction divided by $|\Delta|^2$.
Figure (c) shows the superfluid fraction as a function of the temperature.
In figure (d) we show the superfluid fraction divided by $|\Delta|^2$.
In  figures (a) and (b)  $k_BT/J=0.75$ and in figures (c) and (d) $P=0.10$.
In all the figures all the hopping strengths are same i.e. $J_{\sigma,\alpha}=J_{\sigma',\alpha'}=J$, the average filling fraction $n_{av}=0.50$, $U/J=-6.0$, 
and ${\bf q}=0$.}
\label{fig:fig5}
\end{figure}

Figure~\ref{fig:fig5} (a) shows the superfluid fraction as a function of polarization $P=(n_{\uparrow}-n_{\downarrow})/(n_{\uparrow}+n_{\downarrow})$,
at a constant temperature ($k_BT/J=0.75$).
When the polarization is larger than zero the state is called the Sarma state. One sees, that when the polarization increases the
superfluid fraction decreases. Figure~\ref{fig:fig5} (b) shows the superfluid fraction divided
by $|\Delta|^2$ as a function of the polarization at a constant temperature. When $P$ is about $0.35$ the superfluid fraction divided
by $|\Delta|^2$ drops to zero suddenly. This happens because the pairing gap vanishes at this polarization.  
From figure~\ref{fig:fig5} (b) is seen that the superfluid fraction at a constant temperature is almost proportional to $|\Delta|^2$.
Figure~\ref{fig:fig5} (c) shows the Sarma state superfluid fraction as a function of the temperature,
at a constant polarization ($P=0.10$).
Figure~\ref{fig:fig5} (d) shows the Sarma state superfluid fraction divided by $|\Delta|^2$ ($\rho/|\Delta|^2$). 
We  notice  from figures~\ref{fig:fig5} (b) and (d) that 
$\rho/|\Delta|^2$ is  almost a constant as a function of the temperature and polarization i.e. 
$\rho=c(T,P)|\Delta|^2$, where $c(T,P)$ depends weakly on the temperature and the polarization.

\subsection{Results for FFLO-phase }

When the lattice is cubic and all the hopping strengths are equal the one mode FFLO superfluid fraction is symmetric like the BCS/Sarma superfluid fraction. 
This due the following fact, when one varies the ${\bf q}$, it turns out
that the free-energy is minimized, when the  ${\bf q}$ lies along side the axis i.e. ${\bf q}=q_{\alpha}\hat x_{\alpha}$,  but the system does not favour any of these axis. 

\begin{figure}[!htb]
\centering
\begin{minipage}{0.45\textwidth}
\centering 
\includegraphics[width=\textwidth]{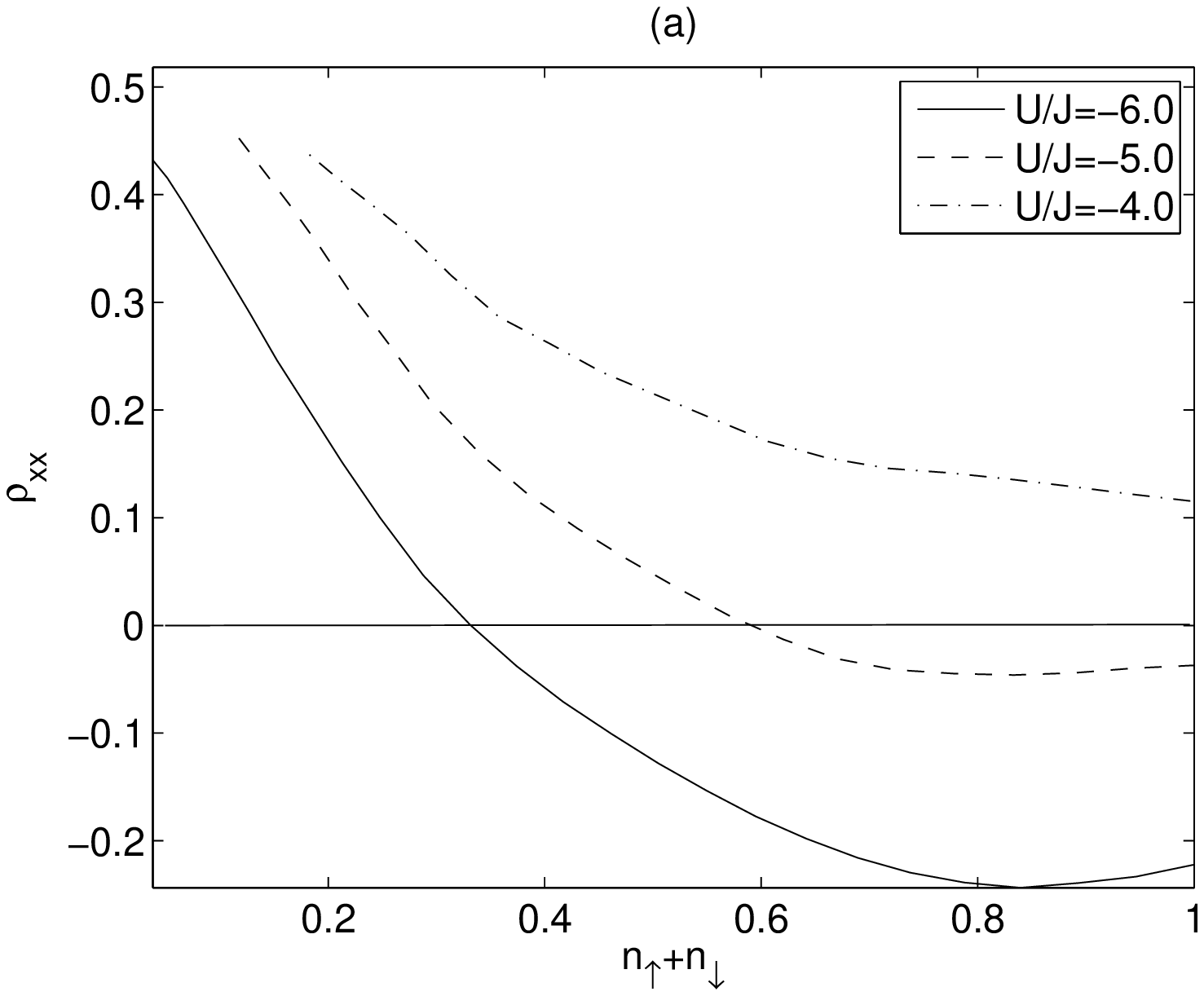}
\end{minipage}
\begin{minipage}{0.45\textwidth}
\centering 
\includegraphics[width=\textwidth]{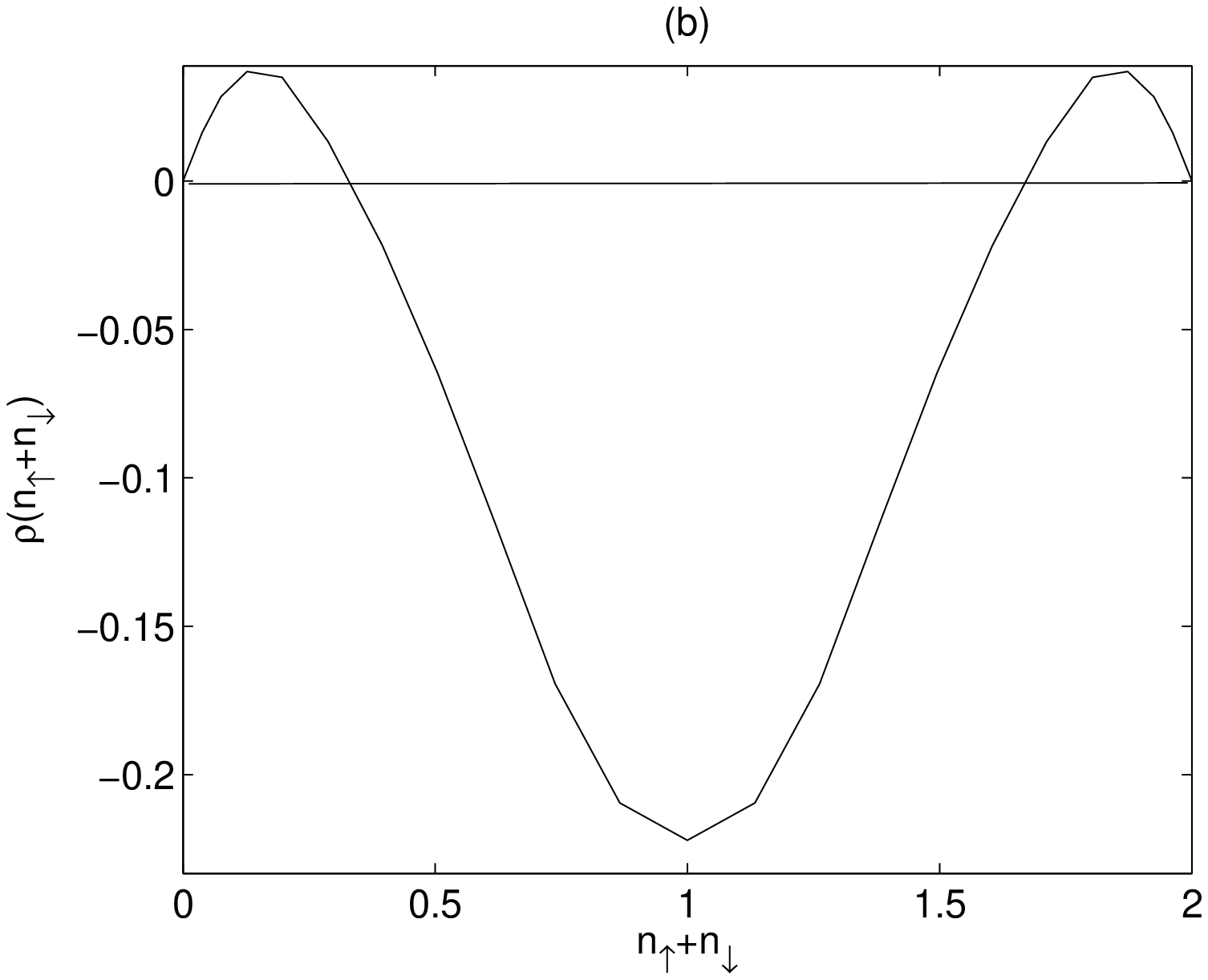}
\end{minipage}
\caption{ The FFLO superfluid fraction at zero temperature.
In figure (a) we show  the FFLO superfluid fraction as a function of the total filling fraction, with three different
interactions at zero temperature.
In figure (b) we show $\rho_{xx}(n_{\uparrow}+n_{\downarrow})$ as a function of the total filling fraction at zero temperature, when
$U/J=-6.0$. In the both figures all the hopping strengths are
same i.e. $J_{\sigma,\alpha}=J_{\sigma',\alpha'}=J$, ${\bf q}=q_x\hat x$, and $P\approx 0.30$.
The horizontal lines in the figures show the value $0$. }
\label{fig:fig6}
\end{figure}

The BCS - and the Sarma state superfluid fractions are almost directly proportional to $|\Delta^2|$ i.e $\rho\sim|\Delta^2|$. 
This changes for the one mode FFLO superfluid fraction as we will now demonstrate.
We show in figure~\ref{fig:fig6} (a)  $\rho_{xx}$-component of the FFLO superfluid fraction as a function of the total filling fraction, with three different
interactions $U/J=-4.0,-5.0,-6.0$, at zero temperature.
figure~\ref{fig:fig6} (b)  shows $\rho_{xx}(n_{\uparrow}+n_{\downarrow})$ as a function of the total filling fraction, when
 $U/J=-6.0$, at zero temperature.
The data which is used in  figure~\ref{fig:fig6} (a) and (b)  have been calculated just above
the Clogston limit~\cite{Clogston1962a}. The Clogston limit is the limit for the chemical potential difference below which
one can find only the BCS type solution when one minimizes the grand potential. The numerical
value of this limit is roughly $\delta\mu\approx \sqrt{2}\Delta_0$, where $\Delta_0$ is the pairing gap at zero temperature when $\delta\mu=0$.
As one can see from  figure~\ref{fig:fig6} (a) when $|U/J|$ increases the superfluid fraction decreases and eventually becomes negative.
One can  also see, that as the total filling fraction increases  the superfluid fraction decreases but does not necessarily become negative.
Negative superfluid fraction implies that the gas is unstable. If superfluid fraction is negative the state around which we expanded is not the ground state of the system, 
and thus the state cannot be stable. In the case where superfluid fraction is negative, the real minimum of the free energy
is beyond our ansatz. 
The real minimum of the system may be a some kind of phase separation or modulating gap state (like the two-mode FFLO state)~\cite{Parish2007a,Chen2007a,Iskin2008a}.

The superfluid density i.e. $\rho_{xx}(n_{\uparrow}+n_{\downarrow})$ is symmetric as a function of the total filling fraction, with the
value $n_{\uparrow}+n_{\downarrow}=1.0$, this is shown in figure~\ref{fig:fig6} (b). This is again due to the particle-hole symmetry of the lattice model.

\begin{figure}[!htb]
\centering
\begin{minipage}{0.45\textwidth}
\centering 
\includegraphics[width=\textwidth]{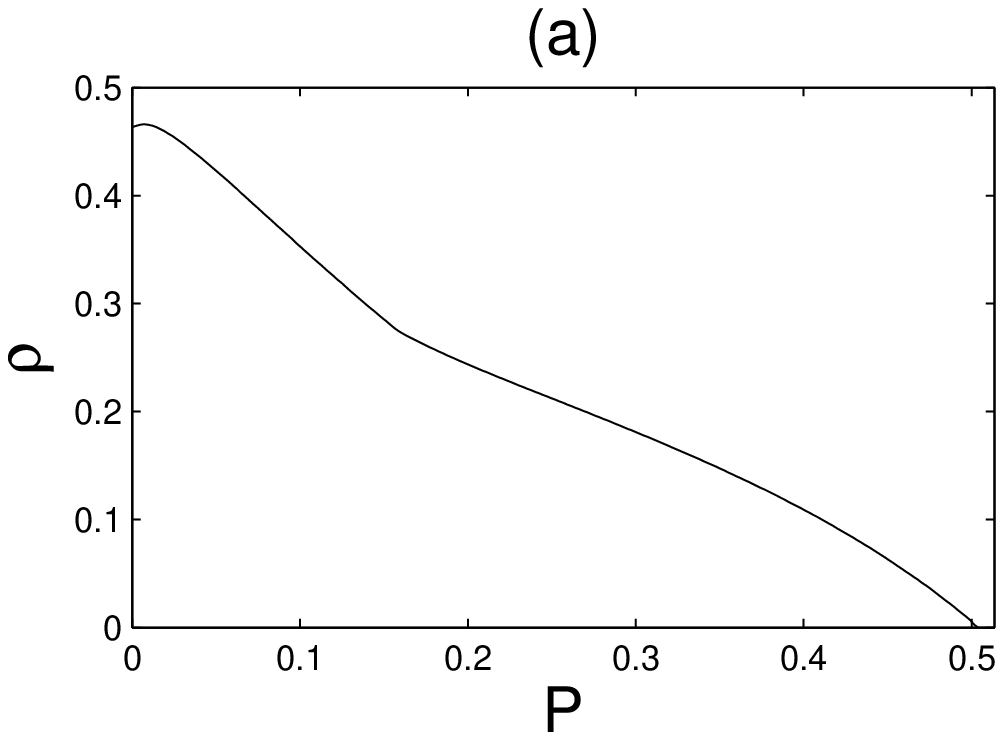}
\end{minipage}
\begin{minipage}{0.45\textwidth}
\centering 
\includegraphics[width=\textwidth]{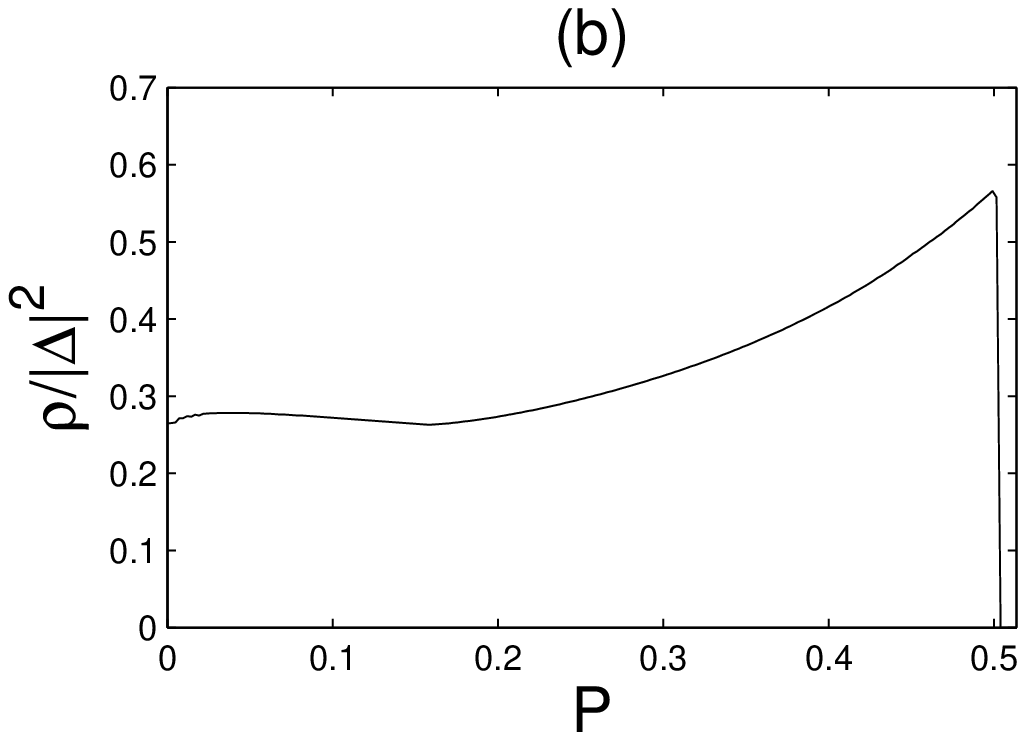}
\end{minipage}
\caption{Figure (a) shows the Sarma- and FFLO phase superfluid fraction as a function of the polarization at constant
temperature.
In figure (b) we show the Sarma- and FFLO phase superfluid fraction divided by $|\Delta|^2$.
The parameters, which were used, are $k_BT/J\approx 0.23$, $U/J\approx -5.1$, and  the average filling fraction is $n_{av}=(n_{\uparrow}+n_{\downarrow})/2=0.2$,
All the hopping strengths were equal. Below $P\approx 0.15$ ${\bf q}=0$ and the phase is the Sarma phase and
above $P\approx 0.15$ ${\bf q}\neq 0$ and the phase is the one mode FFLO phase. 
The sudden drop in figure (b) at $P\approx 0.5$ in indicates that system becomes normal for higher
polarizations. }
\label{fig:fig7}
\end{figure}

Figure~\ref{fig:fig7} (a) shows the superfluid fraction as a function of polarization at a constant temperature ($k_BT/J\approx 0.23$), and figure~\ref{fig:fig7} (b) shows 
the superfluid fraction divided by $|\Delta|^2$ as a function of polarization.
There is a second order phase transition between the Sarma/BCS phase and the FFLO phase, when polarization $P\approx 0.15$, this can be seen
from the bend in the superfluid fraction, clearly. We see also from figure~\ref{fig:fig7} (b) that the FFLO superfluid fraction is not proportional to
$|\Delta|^2$. When one writes $\rho=c(T,P)|\Delta|^2$, the polarization dependence of $c(T,P)$ is very different in the FFLO phase
compared to that in the Sarma phase.

\begin{figure}[!htb]
\centering
\begin{minipage}{0.45\textwidth}
\centering 
\includegraphics[width=\textwidth]{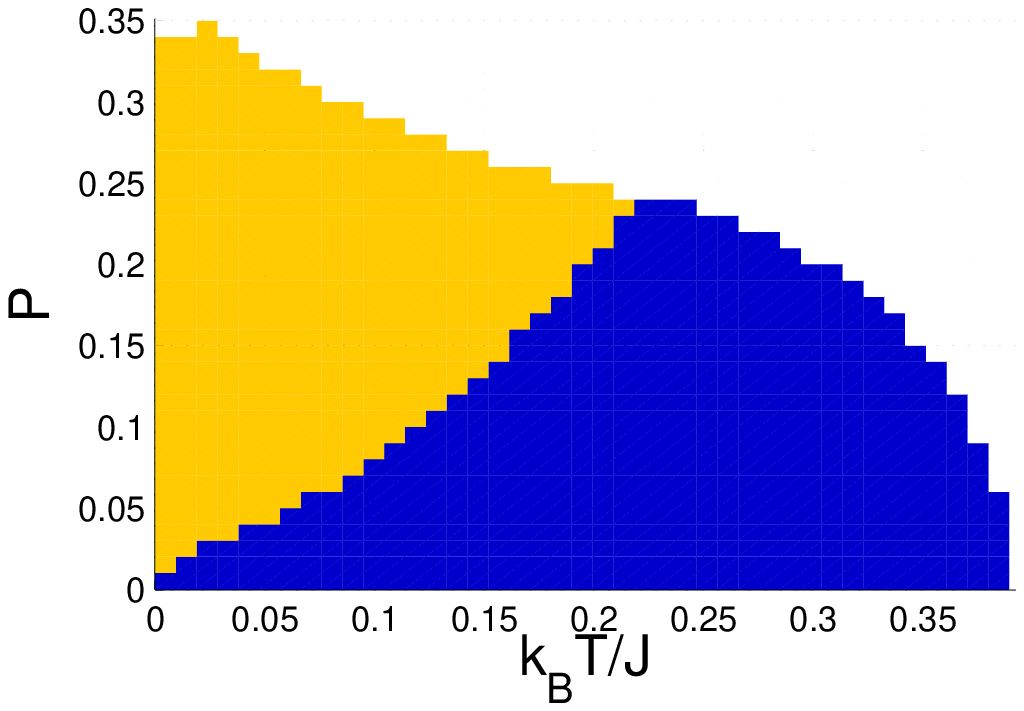}
\end{minipage}
\begin{minipage}{0.45\textwidth}
\centering 
\includegraphics[width=\textwidth]{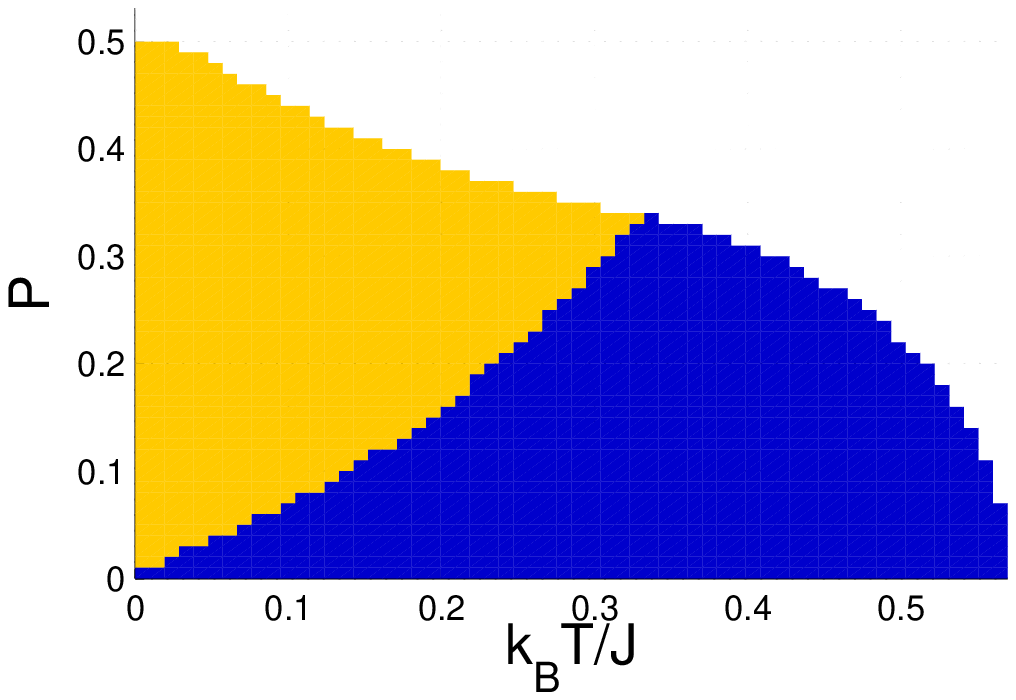}
\end{minipage}
\begin{minipage}{0.45\textwidth}
\centering 
\includegraphics[width=\textwidth]{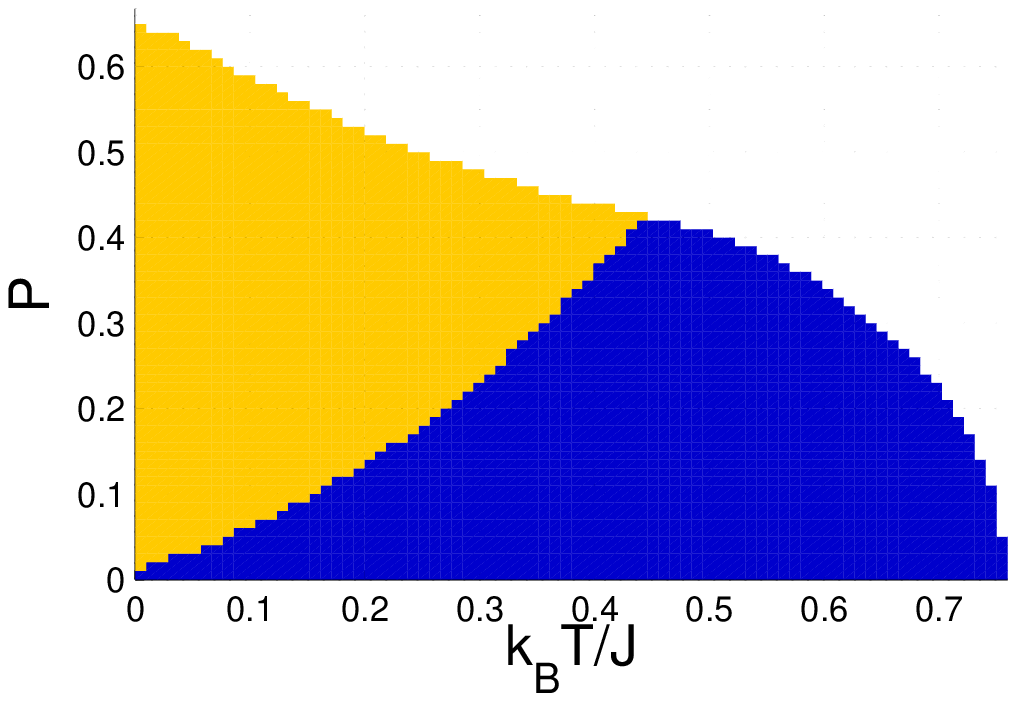}
\end{minipage}
\begin{minipage}{0.45\textwidth}
\centering 
\includegraphics[width=\textwidth]{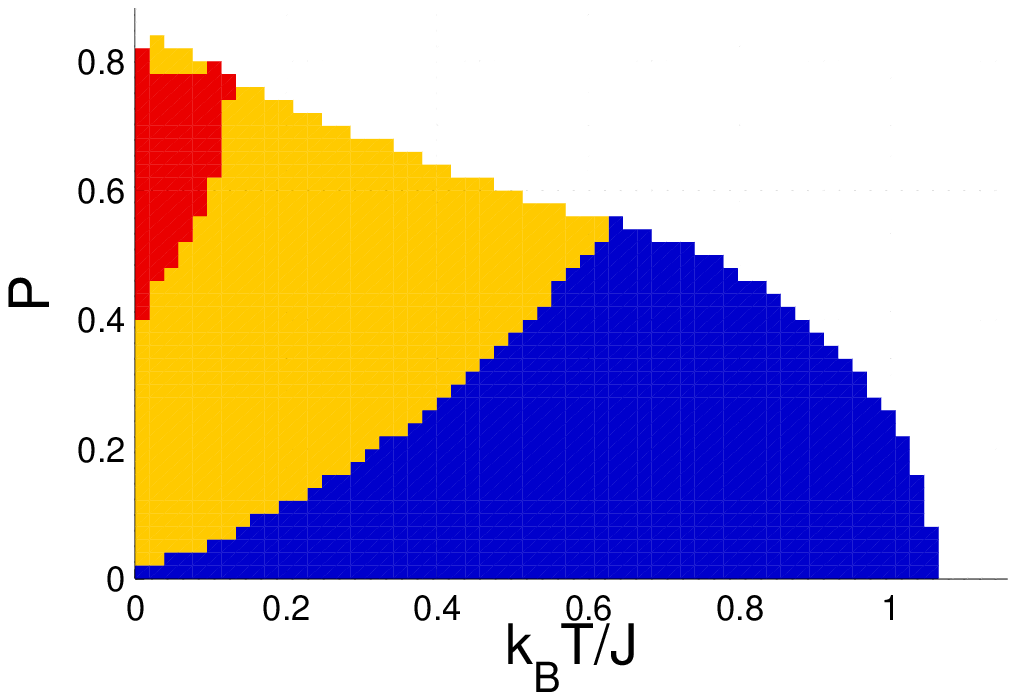}
\end{minipage}
\caption{[Color online] Phase diagrams with four different interaction strengths, when the average filling fraction $n_{av}=0.2$.
The interaction strengths are from up left to bottom right:$U/J\approx -3.7$,$U/J\approx -4.4$, $U/J\approx -5.1$, and
 $U/J\approx -6.3$.
All the hopping strengths are equal.
The color are as follows:BCS/Sarma=blue(black), stable FFLO=yellow(light grey), unstable FFLO=red(dark grey), normal gas=white. }
\label{fig:fig8}
\end{figure}

\begin{figure}[!htb]
\centering
\begin{minipage}{0.45\textwidth}
\centering 
\includegraphics[width=\textwidth]{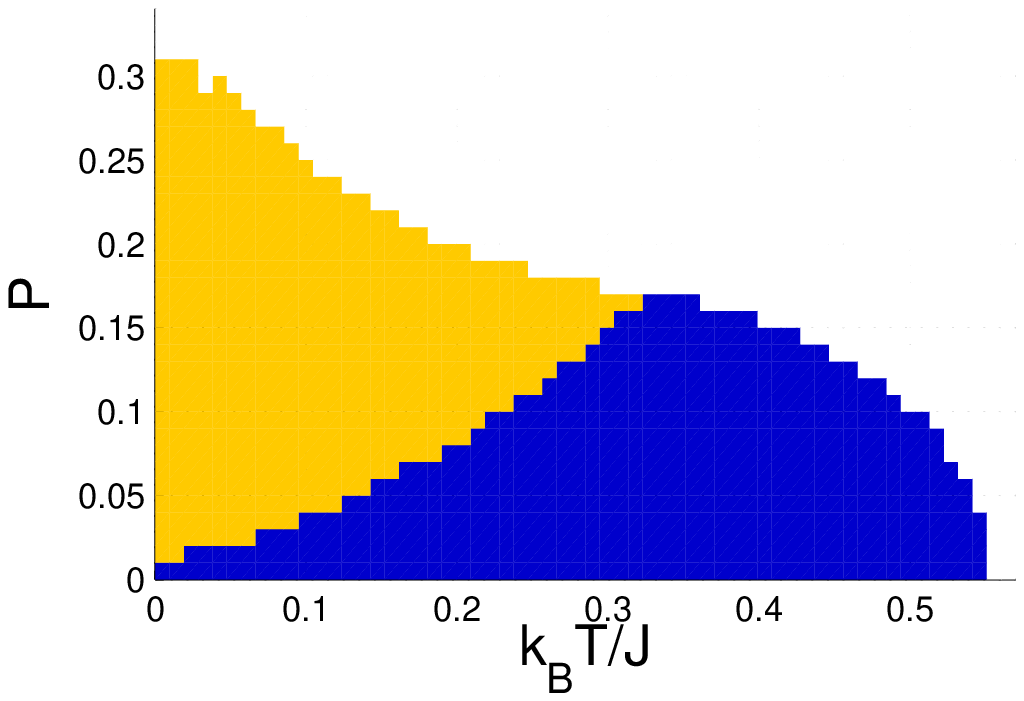}
\end{minipage}
\begin{minipage}{0.45\textwidth}
\centering 
\includegraphics[width=\textwidth]{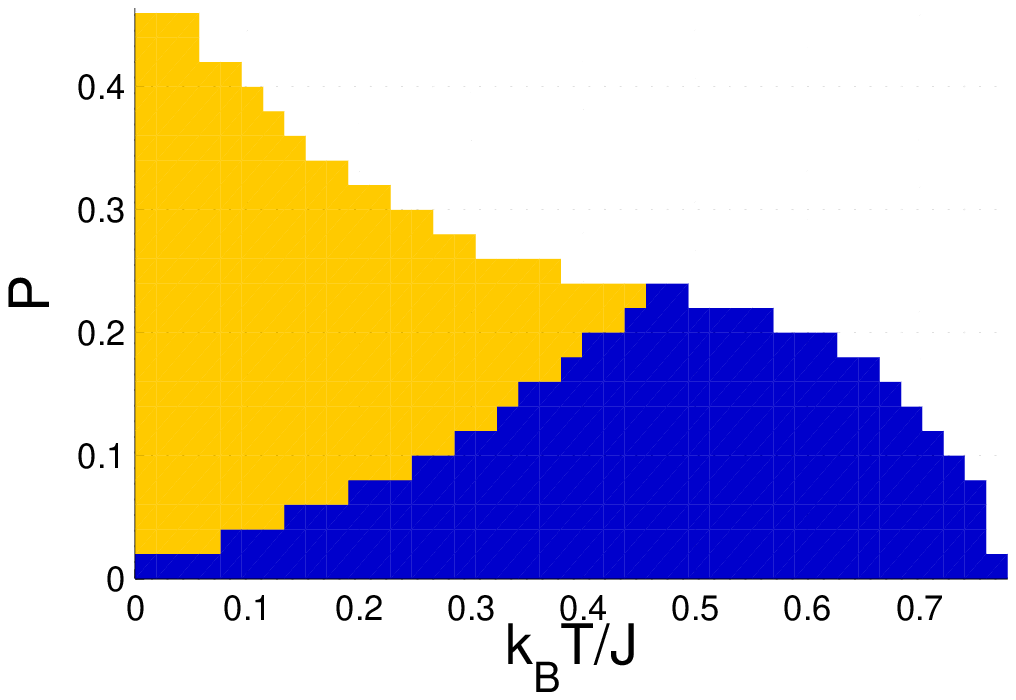}
\end{minipage}
\begin{minipage}{0.45\textwidth}
\centering 
\includegraphics[width=\textwidth]{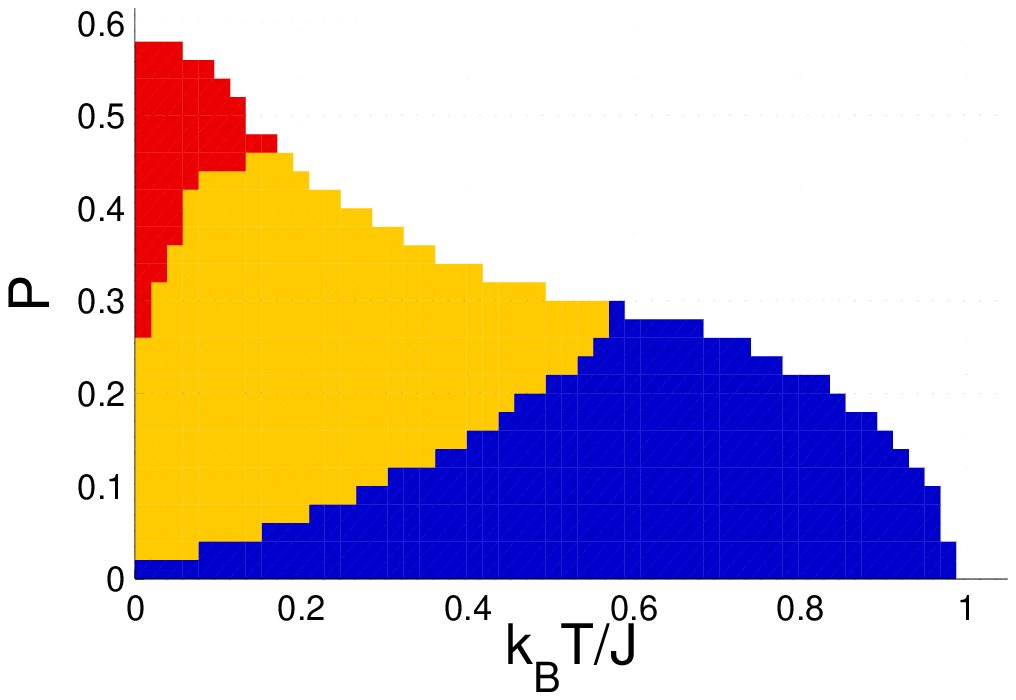}
\end{minipage}
\begin{minipage}{0.45\textwidth}
\centering 
\includegraphics[width=\textwidth]{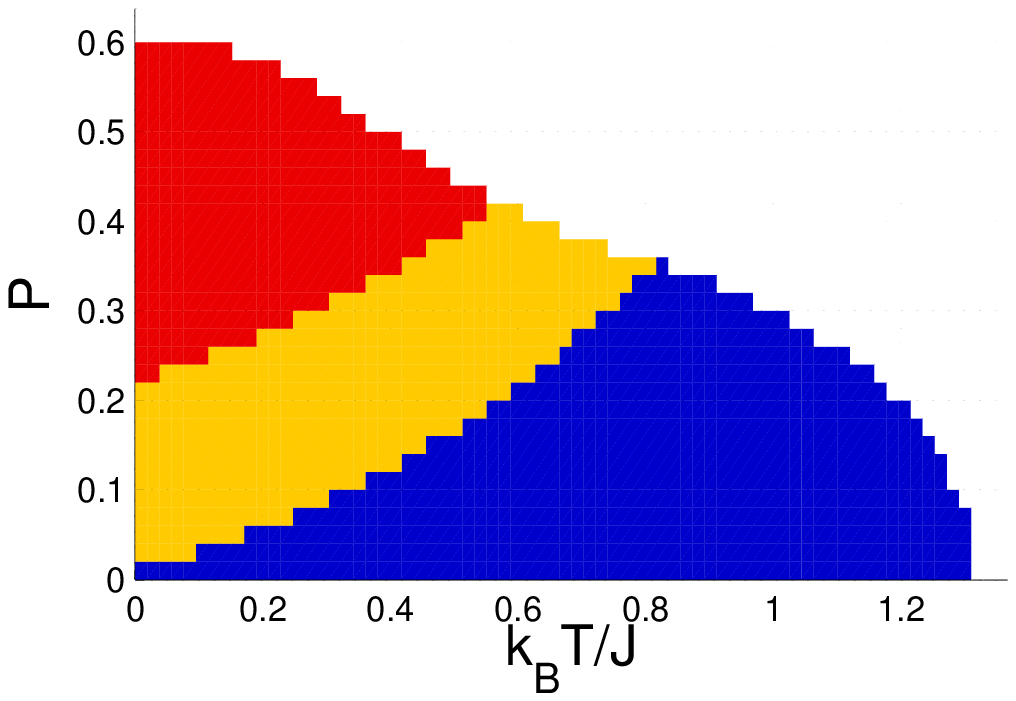}
\end{minipage}
\caption{[Color online] Phase diagrams with four different interaction strengths, when the average filling fraction $n_{av}=0.5$.
The interaction strengths are from up left to bottom right:$U/J\approx -3.7$,$U/J\approx -4.4$, $U/J\approx -5.1$, and
 $U/J\approx -6.3$.
All the hopping strengths are equal.
The color are as follows:BCS/Sarma=blue(black), stable FFLO=yellow(light grey), unstable FFLO=red(dark grey), normal gas=white. }
\label{fig:fig9}
\end{figure}

In figures~\ref{fig:fig8} and~\ref{fig:fig9}  we present phase diagrams with four different interaction strengths, when the average filling fractions are 
$n_{av}=0.2$ and $n_{av}=0.5$, respectively.
The figures show that there can be unstable regions in the FFLO phase. In the unstable FFLO region the superfluid
density is negative, which implies that the phase is dynamically unstable.
However, these instabilities occur only when the interaction is relative high.
By comparing the two figures one notices that the unstable regions increase
when the density increases. Thus when the density decreases the unstable regions eventually
disappear, and this implies that the FFLO phase region is stable without the lattice.
Due to the particle-hole symmetry average filling fraction $0.5$ is the worst case for the stability, but
even when the average density is $0.5$, the one mode FFLO is stable when $|U/J|<4.4$ for any polarization.
The induced interactions might make the FFLO phase more stable, since
these induced interactions make the effective interaction weaker~\cite{Kim2009a}.
References~\cite{Koponen2007a,Koponen2007b} considered also the possibility of
phase separation. However, this work we do not consider phase separation, because
when there is phase separation between the BCS phase and the normal gas, there is
a superfluid gas in a part of the lattice and normal gas in another part of the lattice.
In this case the superfluid density is the standard BCS superfluid density and
the gas is stable. It should be notice that the region of the phase separation does not include the unstable regions of the FFLO phase~\cite{Koponen2007a}.

\section{Conclusions}
\label{sec:conc}

In this paper we have presented, at the mean-field level, the superfluid density of 
the two component Fermi gas in an optical lattice.
We have shown that the BCS superfluid density in optical lattices differs from
the free space results. 
We have also shown that the one-mode FFLO superfluid density differs crucially 
from the BCS/Sarma superfluid density.
In the BCS/Sarma phase the gas is always a stable superfluid,
but in the FFLO phase dynamical instabilities can appear.
However, the FFLO phase is stable, when $|U/J|<4.0$ i.e, on the BCS limit.
Even in the BCS phase the superfluid density can be different in different directions depending on
the lattice structure.

Although the one mode FFLO phase is stable when $|U/J|<4.0$, it is unlikely
that the one mode FFLO is the free energy minimum the system; calculations made
by using the Bogoliubov-de Gennes equations show that the spatially  modulating gap might be more
favorable~\cite{Parish2007a,Chen2007a,Iskin2008a}.
Such states are also, quite likely, more stable than the one mode FFLO phase.

The methods which have used in this paper to calculate the superfluid density, work
in principle for many different systems.
For example, one could use the methods used here to calculate the superfluid
density in the system where another component of the two component Fermi gas occupies the first excited band~\cite{Martikainen2008a}, 
or in the case where an optical lattice is in a trap.
These methods can also been used calculating the superfluid density of modulating gap states.

{\it Acknowledgments}
I would like to thank J.-P. Martikainen and T. K. Koponen for
many illuminating discussions.
This work was supported by Academy of Finland (Project
No. 1110191).

\section{References}

\bibliographystyle{unsrt}
\bibliography{bibli}

\end{document}